\documentclass[preprint,showpacs,preprintnumbers,amsmath,amssymb]{revtex4}

\usepackage{graphicx}
\usepackage{epsfig}
\usepackage{amsmath}
\usepackage{graphics}

\begin{document}

\title{Conversion Efficiencies of Heteronuclear Feshbach Molecules}

\author{Shohei Watabe,$^{1}$ and Tetsuro Nikuni$^{2}$} 
\affiliation{
$^{1}$
Department of Physics, 
Graduate School of Science, 
University of Tokyo, 
3-8-1 Komaba, Meguro-ku, Tokyo, 153-8902, Japan \\
$^{2}$
Department of Physics, Faculty of Science, 
Tokyo University of Science, 
1-3 Kagurazaka, Shinjuku-ku, Tokyo, 
162-8601, Japan}


\begin{abstract}
We study the conversion efficiency of heteronuclear Feshbach molecules in population imbalanced atomic gases 
formed by ramping the magnetic field adiabatically. 
We extend the recent work
[J. E. Williams et al., New J. Phys., {\bf 8}, 150 (2006)] 
on the theory of Feshbach molecule formations 
to various combinations of quantum statistics of each atomic component. 
A simple calculation for a harmonically trapped ideal gas 
is in good agreement with the recent experiment 
[S. B. Papp and C. E. Wieman, 
Phys. Rev. Lett., {\bf 97}, 180404 (2006)] 
{\it without} any fitting parameters. 
We also give the conversion efficiency as an explicit function of initial peak phase space density of 
the majority species for population imbalanced gases. 
In the low-density region where Bose-Einstein condensation does not appear, 
the conversion efficiency is a monotonic function of the initial peak phase space density, 
but independent of statistics of a minority component. 
The quantum statistics of majority atoms 
has a significant effect on the conversion efficiency. 
In addition, Bose-Einstein condensation of an atomic component 
is the key element determining the maximum conversion efficiency. 
\end{abstract}


\maketitle

\section{Introduction}
There have been a number of efforts to produce molecules in the ultracold atomic gases.
Making use of the Feshbach resonance, 
{\it homonuclear} molecular conversion efficiencies 
in atomic gases of bosonic $^{85}$Rb atoms and fermionic $^{40}$K atoms were investigated 
by Hodby {\it et al.}~\cite{hodby:2005}. 
They found that the conversion efficiencies do not reach 100$\%$ 
even for an adiabatic sweep of the magnetic field, 
but depend on the peak phase space density of initially prepared pure atomic gases.

The ultracold diatomic {\it heteronuclear} molecule is now drawing attention. 
For example, there is an expectation of realizing quantum computation 
by making use of the electric dipole moments of heteronuclear molecules~\cite{demille:2002}. 
So far several basic experiments have been performed in order to realize this idea. 
Heteronuclear Feshbach resonances 
have been investigated 
in collisions between 
bosonic $^{23}$Na atoms and fermionic $^{6}$Li atoms~\cite{stan:2004}, 
bosonic $^{87}$Rb atoms and fermionic $^{40}$K atoms~\cite{inouye:2004}, 
and 
bosonic $^{87}$Rb atoms and bosonic $^{85}$Rb atoms~\cite{papp:2006}. 
The adiabatic conversion efficiency of heteronuclear Feshbach molecules 
composed of $^{87}$Rb and $^{85}$Rb in population imbalanced gas 
have been investigated in detail~\cite{papp:2006}. 
Fermionic heteronuclear Feshbach molecules 
via rf association 
are investigated in a 3D optical lattice~\cite{COspelkaus:2006}. 


In order to analyze the experimental data for the molecular conversion efficiency, 
Hodby {\it et al.}~\cite{hodby:2005} devised a stochastic phase space sampling (SPSS) model 
based on the assumption 
that only a pair of atoms sufficiently close in phase space 
can form a molecule through the adiabatic process. 
With a single fitting parameter $\gamma$ (a cutoff-radius in phase space), 
which determines how close two atoms should be in the phase space to form a molecule, 
the phenomenological SPSS model predicted 
that the molecular conversion efficiency 
is a universal function of the initial peak phase space density. 
The prediction from the SPSS model agreed notably well 
with the experimental data~\cite{hodby:2005}. 
The SPSS model was also used to analyze the 
conversion efficiencies $^{87}$Rb-$^{85}$Rb Feshbach molecules~\cite{papp:2006}, 
which was in agreement with the experimental data.  

On the other hand, 
Williams {\it et al.}~\cite{williams:2006} provided the theory of Feshbach molecule formation based on 
the coupled atom-molecule Boltzmann equation. 
They derived from the first principles 
the universal functional form for the conversion efficiency without any fitting parameters. 
This approach is appropriate 
for the regime $t_{\rm ramp} \gg \tau_{\rm col}$, 
where multiple collisions occur. 
Here, $t_{\rm ramp}$ is an upper bound on the ramp time 
and $\tau_{\rm col}$ 
is the average time between collisions. 

The coupled atom-molecule Boltzmann approach 
provided a result 
that the exchange of atoms and molecules ceases 
when the conservation laws of momentum and energy 
via the atom-molecule conversion cannot be satisfied~\cite{williams:2006,williams:2004}. 
When we neglect the energy shift by the real part of the self-energy, 
these conservation laws are not satisfied for negative detunings. 
The detuning is the difference between 
an energy of the atomic dissociation continuum 
and 
the lowest single particle energy of the molecule, 
which is controlled by the magnetic field due to the Zeeman shift. 

According to Williams {\it et al.}~\cite{williams:2006}, 
the conversion mechanism can be described as follows. 
When the starting point is the positive detuning far from the resonance 
with pure atomic gas, 
Feshbach molecules are formed and decay 
as the detuning is ramped in the region $\delta > 0$. 
When the detuning crosses zero and becomes negative, 
the molecule formation effectively halts, 
which gives rise to the saturation of molecule production 
because the momentum and energy conservation laws are not satisfied. 
If the density is high and 
the 3-body recombination to form the molecule is effective, 
the saturation of molecule population would not appear, 
and the molecule formation would take place even in negative detunings. 
However, 
the saturation of the molecule population 
against the magnetic field is observed 
in many experiments~\cite{hodby:2005,regal:2003,strecker:2003,regal:2004:jan,
regal:2004:feb,zwierlein:2004,papp:2006}. 
Therefore, one is allowed to assume 
that 
the sweep speed of the magnetic field is fast with respect to the time scale for 
3-body collisions and only 
the atom-molecule resonant collisions 
play a role in producing the molecule in these experiments. 

On the other hand, the saturation against the inverse sweep rate is reported 
in experiments~\cite{hodby:2005,regal:2003,strecker:2003}. 
The saturation of molecular population with slow sweep 
means that the detailed balance is realized. 
Using the kinetic theory, 
Williams {\it et al.}~\cite{williams:2006} 
signified that the atom-molecule resonant collisions alone can give rise 
to thermal equilibrium as well as chemical equilibrium~\cite{williams:2004}. 
Considering the assumption 
that a sweep is slow with respect to 2-body process 
but it is fast with respect to the time scale for 3-body recombinations, 
they concluded that the detailed balance is only realized 
for positive detunings only through atom-molecule resonant collisions. 
We note that the production rate of entropy vanishes 
as long as the slow sweep of the magnetic field 
keeps the system in local equilibrium. 
Therefore, when the sweep is so slow that the molecular conversion rate gets saturated, 
the process may be regarded as adiabatic for positive detunings. 
For negative detunings, the chemical equilibrium is no longer maintained. 

As a result, 
the conversion efficiency with adiabatic sweep 
is determined by the molecular fraction at zero detuning, 
which is a function of the physical quantity of initial state 
connected through the equal entropy. 
This mechanism does not require any fitting parameters. 
According to this mechanism, 
the conversion efficiencies plotted against the initial peak phase 
space density~\cite{williams:2006} 
agreed well qualitatively with the experiment~\cite{hodby:2005}, 
even though using an ideal gas mixture model. 
Although these saturation mechanisms are derived 
above the superfluid transition temperature, 
these mechanisms are based on the momentum and energy conversion laws 
so that they are assumed to be quite general. 
The principle given in Ref.~\cite{williams:2006} 
was extended to the superfluid phase of a Fermi gas 
below the transition temperature~\cite{watabe:2006}. 
Conversion efficiency 
against the initial temperature~\cite{watabe:2006} 
including the resonant interaction 
agreed well qualitatively with the experiment~\cite{hodby:2005}. 

Within the classical gas approximation, 
the explicit formula for the conversion efficiency 
was given 
as a function of the initial peak phase space density~\cite{williams:2006}, 
which shows a qualitative agreement with the experiment~\cite{hodby:2005}, 
consistent with the SPSS model.

In this paper, we study 
the conversion efficiency of the heteronuclear Feshbach molecule 
in population imbalanced gases, 
extending the theoretical model provided by Williams {\it et al.}~\cite{williams:2006}. 
We will show that our calculation agrees with the experimental result 
by Papp and Wieman~\cite{papp:2006} {\it without} any fitting parameters. 
An exquisite point of this theoretical model 
is that one does not need any fitting parameters 
so that one can make quantitative prediction 
for the conversion efficiencies 
which have not been measured yet. 
Thus, we investigate various combinations of quantum statistics of atoms. 
All combinations studied in this paper are collected on the table below. 
Symbols ``${\rm B}$'' and ``${\rm F}$'' represent the quantum statistics of 
boson and fermion, and 
symbols ``$_{>}$'', ``$_{<}$'' and ``$_{\rm m}$'' represent 
the majority atomic component, 
the minority atomic component, 
and the molecular component, 
\begin{eqnarray}
\left \{
\begin{array}{lll}
(1)&&{\rm B}_{>}+{\rm B}_{<}\leftrightarrow {\rm B}_{\rm m},
\\
(2)&&{\rm B}_{>}+{\rm F}_{<}\leftrightarrow {\rm F}_{\rm m},
\\
(3)&&{\rm F}_{>}+{\rm B}_{<}\leftrightarrow {\rm F}_{\rm m},
\\
(4)&&{\rm F}_{>}+{\rm F}_{<}\leftrightarrow {\rm B}_{\rm m}.
\end{array}
\right.
\end{eqnarray} 
We also give the conversion efficiency 
as an explicit function of the initial peak phase space density of 
the majority atomic component. 
We will show that in the low-density region 
where Bose-Einstein condensation does not appear 
the conversion efficiency is a monotonic function of the initial peak phase space density, 
but independent of statistics of the minority component. 

For {\it homonuclear} Feshbach molecules, 
the trend of conversion efficiencies calculated 
by Williams {\it et al.}~\cite{williams:2006} 
was qualitatively consistent with the SPSS model 
within the range plotted in Refs.~\cite{hodby:2005} and~\cite{williams:2006}. 
However, 
we will show that the model by Williams {\it et al.}~\cite{williams:2006} 
leads to completely different results from 
the phenomenological SPSS model at low temperatures, 
when applied to {\it heteronuclear} Feshbach molecules. 
In the case of bosonic {\it homonuclear} Feshbach molecules 
composed of bosonic atoms, 
the trend of conversion efficiency is also expected to have a completely different result 
from the SPSS model in the presence of Bose-Einstein condensation. 
Although the SPSS model has shown agreement with the experiments so far, 
the problem with this model 
with regard to heteronuclear Feshbach molecules 
is that the atoms of different species are disposed 
in the same phase space, 
although they should be essentially in own phase spaces. 

\section{Equilibrium theory}
In this Section, we present the equilibrium theory 
for the ideal gas mixture composed of two species of atoms 
and heteronuclear Feshbach molecules, 
assuming that atomic populations are imbalanced. 

We denote the numbers of majority atoms, minority atoms and molecules 
as $N_{>}$, $N_{<}$ and $N_{\rm m}$, respectively. 
The following two constraints should be satisfied in these systems. 
The first constraint is the conservation of the total number of particles: 
\begin{eqnarray}
N_{\rm tot}&=&N_{>}+N_{<}+2N_{\rm m}.
\label{eq2}
\end{eqnarray}
The second constraint is the conservation of the ratio: 
\begin{eqnarray}
\frac{N_{<}+N_{\rm m}}{N_{>}+N_{\rm m}}
= \alpha, 
\label{eq3}
\end{eqnarray}
where $\alpha$ is determined 
by initial ratio of $N_{<,{\rm ini}}$ and $N_{>,{\rm ini}}$ 
(where there is no molecular component), 
{\it i.e.}, 
$N_{<,{\rm ini}}/N_{>,{\rm ini}} \equiv \alpha$. 
By definition of the majority and minority components, 
one has $\alpha \leq 1$. 
The second constraint can also be written as 
\begin{eqnarray}
N_{<}-\alpha N_{>}+(1-\alpha)N_{\rm m}=0. 
\label{eq4}
\end{eqnarray}

In order to impose two constraints in Eqs. (\ref{eq2}) and (\ref{eq4}) described above, 
we deal with the grand canonical Hamiltonian 
with introducing two Lagrange multipliers $\mu_{1}$ and $\mu_{2}$. 
The partition function ${\Xi}$ for this grand canonical Hamiltonian 
is defined by 
\begin{eqnarray}
\Xi \equiv {\rm Tr} \left [
\exp{\left ( 
-\beta \left \{
\hat{H}_{0}
-\mu_{1}\left ( \hat{N}_{>} + \hat{N}_{<} + 2\hat{N}_{\rm m} \right )
-\mu_{2}\left [ \hat{N}_{<} - \alpha\hat{N}_{>} + (1-\alpha)\hat{N}_{\rm m} \right ]
\right \}
\right )}
\right ], 
\end{eqnarray}
where 
$\beta = 1/k_{\rm B}T$ with $k_{\rm B}$ and $T$ 
being the Boltzmann's constant and temperature. 
The Hamiltonian $\hat{H}_{0}$ for the ideal gas mixtures is given by 
\begin{eqnarray}
\hat{H}_{0} \equiv 
\sum\limits_{i}\varepsilon_{i}^{>}\hat{n}_{i}^{>}
+\sum\limits_{j}\varepsilon_{j}^{<}\hat{n}_{j}^{<}
+\sum\limits_{k}(\varepsilon_{k}^{\rm m}+\delta)\hat{n}_{k}^{\rm m}, 
\end{eqnarray}
where $\varepsilon_{i}^{>}$, $\varepsilon_{j}^{<}$ and $\varepsilon_{k}^{\rm m}$
are the single-particle energies, 
and $\hat{n}_{i}^{>}$, $\hat{n}_{j}^{<}$, and $\hat{n}_{k}^{\rm m}$ are the number operators. 
The lowest energy of the molecule is displaced 
relative to that of the atomic dissociation continuum by the detuning $\delta$, 
which is controlled by the magnetic field. 
Operators of atomic and molecular populations are given by 
\begin{eqnarray}
\hat{N}_{>} = \sum\limits_{i}\hat{n}_{i}^{>}, \qquad 
\hat{N}_{<} = \sum\limits_{j}\hat{n}_{j}^{<}, \qquad
\hat{N}_{\rm m} = \sum\limits_{k}\hat{n}_{k}^{\rm m}. 
\end{eqnarray}
The thermodynamic potential $\Omega$ is defined by 
$\Omega \equiv - k_{\rm B}T \ln{\Xi}$. 
We derive two equations representing two constraints 
from the thermodynamic potential $\Omega$. 
The total number of particles is given by 
$N_{\rm tot} = - \partial \Omega/\partial \mu_{1}$, which leads to 
\begin{eqnarray}
N_{\rm tot}&=&
\sum\limits_{i}
\frac{1}{z_{>}^{-1}e^{\beta\varepsilon_{i}^{>}}\mp 1}
+
\sum\limits_{j}
\frac{1}{z_{<}^{-1}
e^{\beta\varepsilon_{j}^{<}}\mp 1}
+
2\sum\limits_{k}
\frac{1}{z_{\rm m}^{-1}
e^{\beta\varepsilon_{k}^{\rm m}}\mp 1}. 
\label{eq5}
\end{eqnarray}
The equation representing the second constraint is 
given by 
$0 = - \partial \Omega/\partial \mu_{2}$, 
which leads to 
\begin{eqnarray}
0&=&
\sum\limits_{j}
\frac{1}{z_{<}^{-1}e^{\beta\varepsilon_{j}^{<}}\mp 1}
-\alpha\sum\limits_{i}
\frac{1}{z_{>}^{-1}e^{\beta\varepsilon_{i}^{>}}\mp1}
+(1-\alpha)
\sum\limits_{k}
\frac{1}{z_{\rm m}^{-1}
e^{\beta\varepsilon_{k}^{\rm m}}\mp 1}. 
\label{eq6}
\end{eqnarray}
In Eqs.~(\ref{eq5}) and (\ref{eq6}), 
the negative sign is for bosons and the positive sign is for fermions. 
The fugacities of the majority atoms, the minority atoms and the heteronuclear molecules 
are defined by 
\begin{eqnarray}
{\it z}_{>} \equiv e^{(\mu_{1}-\alpha\mu_{2})/k_{\rm B}T},
&
{\it z}_{<}  \equiv e^{(\mu_{1}+\mu_{2})/k_{\rm B}T},
&
{\it z}_{\rm m}  \equiv 
e^{[2\mu_{1}+(1-\alpha)\mu_{2}-\delta]/k_{\rm B}T}.
\label{fugacities}
\end{eqnarray} 
These fugacities are related to each other through 
\begin{eqnarray}
z_{>}z_{<} = z_{\rm m}e^{\delta/k_{\rm B}T}. 
\label{eq8}
\end{eqnarray} 
The relation at zero detuning $\delta=0$ 
given by $z_{>}z_{<} = z_{\rm m}$ will play an important role 
in considering the molecular conversion efficiency.

We consider a gas confined in an anisotropic harmonic trap 
with frequencies $\{ \omega_{x}^{i}$, $\omega_{y}^{{i}}, \omega_{z}^{{i}}\}$, 
where $i$ represents the component indexes $\{ >, <, {\rm m}\}$. 
For later use, we define the ratio of the trap frequencies 
$\gamma_{<} \equiv \bar{\omega}_{>}/\bar{\omega}_{<}$ and 
$\gamma_{\rm m} \equiv \bar{\omega}_{>}/\bar{\omega}_{\rm m}$, 
where 
$\bar{\omega}_{i}\equiv
(\omega_{x}^{i} \omega_{y}^{i} \omega_{z}^{i})^{1/3}$. 
In optical dipole traps, the trap frequencies are determined 
by the polarizability and the particle masses~\cite{Morales}.

As in Ref.~\cite{williams2004adiabatic}, 
we assume $k_{\rm B}T \gg \hbar \bar{\omega}_{i}$ 
and thus use the semi-classical formulation, 
which replaces the sum over discrete states in Eqs. (\ref{eq5}) and (\ref{eq6}) 
by an integral.  
The density of states $\rho_{i}(\varepsilon )$ in a harmonic trap is given by 
$\rho_{i}(\varepsilon) = 
\varepsilon^{2}/[2(\hbar \bar{\omega}_{i})^{3}]$. 
The noncondensed population $\tilde{N}_{\rm B}$ for bosons and 
the population $N_{\rm F}$ for fermions are represented 
by the Bose integral $\mathcal{G}_{n}(z)$ and 
the Fermi integral $\mathcal{F}_{n}(z)$ respectively, 
which are given by~\cite{williams2004adiabatic}, 
\begin{eqnarray}
\tilde{N}_{\rm B} = \left (\frac{k_{\rm B}T}{\hbar\bar{\omega}}\right )^{3}
\mathcal{G}_{3}({\it z}), 
&&
N_{\rm F} = \left (\frac{k_{\rm B}T}{\hbar\bar{\omega}}\right )^{3}
\mathcal{F}_{3}({\it z}), 
\end{eqnarray} 
where $z$ is a fugacity, 
and $\bar{\omega}$ is the geometric average of the trap frequency 
$\bar{\omega} = (\omega_{x}\omega_{y}\omega_{z})^{1/3}$. 
The Bose and Fermi integrals are 
defined by 
\begin{eqnarray}
{\mathcal G}_{n}({\it z}) 
\equiv
\frac{1}{\Gamma(n)}
\int_{0}^{\infty}
\frac{x^{n-1}dx}{{\it z}^{-1}e^{x}-1}, 
&&
{\mathcal F}_{n}({\it z}) 
\equiv
\frac{1}{\Gamma(n)}
\int_{0}^{\infty}
\frac{x^{n-1}dx}{{\it z}^{-1}e^{x}+1}, 
\end{eqnarray}
where $\Gamma(n)$ is the gamma function. 
The Bose function is equal to the Riemann-zeta function $\zeta(n)$ 
when $z = 1$, i.e., $\mathcal{G}_{n}(z = 1) = \zeta(n)$. 

The entropy of noncondensed bosons $\tilde{S}_{\rm B}$ and 
the entropy of fermions $S_{\rm F}$ are given by~\cite{williams2004adiabatic} 
\begin{eqnarray}
\tilde{S}_{\rm B}(z) = k_{\rm B} \tilde{N}_{\rm B} 
\left[4\frac{\mathcal{G}_{4}({\it z})}{\mathcal{G}_{3}({\it z})}
-\ln{{\it z}}\right],
&&
S_{\rm F}(z) =k_{\rm B} N_{\rm F}
\left[4\frac{\mathcal{F}_{4}({\it z})}{\mathcal{F}_{3}({\it z})}
-\ln{{\it z}}\right]. 
\label{entropyBF}
\end{eqnarray} 
In the low temperature limit, 
they are approximated by 
\begin{eqnarray}
\tilde{S}_{\rm B}\approx 4k_{\rm B}
\left(\frac{k_{\rm B}T}{\hbar\bar{\omega}}\right)^{3}\zeta(4), 
&&
S_{\rm F} \approx k_{\rm B} \pi^{2} 
\left( \frac{N_{\rm F}^{2}}{6} \right)^{1/3}
\frac{k_{\rm B}T}{\hbar\bar{\omega}}.
\end{eqnarray}

We assume that the initially prepared state has a large and positive detuning 
so that no molecules exist in the initial state. 
The initial populations of atoms satisfying two constraints 
in Eqs. (\ref{eq2}) and (\ref{eq4}) are given by 
\begin{eqnarray}
N_{>,{\rm ini}}=\frac{1}{1+\alpha}N_{\rm tot}, &&
N_{<,{\rm ini}}=\frac{\alpha}{1+\alpha}N_{\rm tot}. 
\end{eqnarray}

According to the model for the molecular conversion efficiency 
proposed in Ref.~\cite{williams:2006}, 
we define the molecular conversion efficiency $\chi_{0}$ 
as a fraction of a molecular population $N_{\rm m}$ 
at zero detuning $\delta = 0$, 
given by 
\begin{eqnarray}
\chi_{0} = \frac{N_{\rm m}(\delta=0)}{N_{<,{\rm ini}}}
=\frac{N_{\rm m}(\delta=0)}{N_{\rm tot}}\frac{1+\alpha}{\alpha}. 
\end{eqnarray}

\section{Bosonic heteronuclear molecules composed of bosonic atoms   
 \label{SecBBB}}
In this Section, 
we consider the first case: $\{ {\rm B}_{>}+{\rm B}_{<}\leftrightarrow {\rm B}_{\rm m}\}$. 
To make contact with the experiment of Ref.~\cite{papp:2006}, 
we relabel two atomic components $^{87}$Rb and $^{85}$Rb 
as $87$ and $85$ respectively, 
where $^{87}$Rb is a majority component and $^{85}$Rb is a minority component. 
The value of $\alpha$ is chosen from experimental data; 
we adopt $\alpha = 2/15$, 
assuming $N_{85,{\rm ini}}=40,000$ and $N_{87,{\rm ini}}=300,000$ 
reported by Papp {\it et al.}~\cite{papp:2006} 
to be typically produced.

We note that the two constraints in Eqs. (\ref{eq2}) and (\ref{eq4}) are given as 
\begin{eqnarray}
\left \{
\begin{array}{lll}
N_{\rm tot}&=& 
\left[N_{\rm c}^{87}
+
\displaystyle{
\left(\frac{k_{\rm B}T}{\hbar\bar{\omega}_{87}}\right)^{3}
}
\mathcal{G}_{3}({\it z}_{87})\right ] 
+ \left[N_{\rm c}^{85} + 
\displaystyle{
\left(\frac{k_{\rm B}T}{\hbar\bar{\omega}_{85}}\right)^{3}
}
\mathcal{G}_{3}({\it z}_{85})\right] 
\\
\\
&& 
\qquad \qquad 
\qquad \qquad 
\qquad \qquad \qquad  
+ 2 \left[N_{\rm c}^{\rm m}+
\displaystyle{
\left(\frac{k_{\rm B}T}{\hbar\bar{\omega}_{\rm m}}\right)^{3}
}
\mathcal{G}_{3}({\it z}_{\rm m})\right],
\\
\\
0&=& 
\left[N_{\rm c}^{85} + 
\displaystyle{
\left(\frac{k_{\rm B}T}{\hbar\bar{\omega}_{85}}\right)^{3}
}
\mathcal{G}_{3}({\it z}_{85})\right]
-\alpha\left[N_{\rm c}^{87}+
\displaystyle{
\left(\frac{k_{\rm B}T}{\hbar\bar{\omega}_{87}}\right)^{3}
}
\mathcal{G}_{3}({\it z}_{87})\right] 
\\
\\
&& 
\qquad \qquad 
\qquad \qquad 
\qquad \qquad
+ (1-\alpha)\left[N_{\rm c}^{\rm m}+
\displaystyle{
\left(\frac{k_{\rm B}T}{\hbar\bar{\omega}_{\rm m}}\right)^{3}
}
\mathcal{G}_{3}({\it z}_{\rm m})\right], 
\end{array}
\right.
\end{eqnarray}
where $N_{\rm c}^{87}$, $N_{\rm c}^{85}$ and $N_{\rm c}^{\rm m}$ 
are condensed populations of the respective components.

According to the experiment~\cite{papp:2006}, 
$^{87}$Rb is a majority component, 
compared with $^{85}$Rb and the heteronuclear molecule. 
This makes the Bose-Einstein condensation (BEC) transition temperature 
of $^{87}$Rb atoms, 
which we define as $T_{\rm c}^{87}$, higher than that of $^{85}$Rb atoms 
and heteronuclear molecules. 
The condition for BEC of $^{87}$Rb component is given by 
\begin{eqnarray}
\mu_{1} = \alpha\mu_{2}. 
\end{eqnarray}
We note that the fugacities ${\it z}_{85}$ and ${\it z}_{\rm m}$ 
given in Eq.~(\ref{fugacities}) are always less than 1, 
because the distribution functions must be positive. 
Below $T_{\rm c}^{87}$, these conditions reduce to 
\begin{eqnarray}
\left \{
\begin{array}{ccl}
(1+\alpha)\mu_{2}&\leq&0, 
\\
(1+\alpha)\mu_{2}-\delta&\leq&0, 
\end{array}
\right.
\label{eq17}
\end{eqnarray}
where we have used the condition $\mu_{1} = \alpha\mu_{2}$.

We shall determine the phase diagram 
in $\delta$-$T$ plane. 
Below $T_{\rm c}^{87}$, 
there are some possible cases 
where $^{85}$Rb atoms or heteronuclear molecules, or both could be condensed. 
We consider three separate cases; 
$\delta > 0$, $\delta < 0$ and 
$\delta = 0$, 
because populations of bosonic atoms $^{87}$Rb and $^{85}$Rb 
and bosonic heteronuclear molecules are 
sensitive to the sign of the detuning $\delta$.  

We first consider the case $\delta > 0$. 
According to the inequalities in Eq.~(\ref{eq17}) and $\delta > 0$, 
we find that 
$\delta - (1+\alpha)\mu_{2} $ 
cannot reach $0$. 
In other words, heteronuclear molecule can never be condensed ($N_{\rm c}^{\rm m}=0$). 
On the other hand, BEC of $^{85}$Rb atoms takes place 
when $\mu_{2} = 0$. 
This is reasonable because the lowest energy 
of heteronuclear molecules is higher than 
that of $^{85}$Rb atoms. 
We determine $T_{\rm c}^{+}$ 
below which BEC of $^{85}$Rb atoms takes place from the following equation, 
\begin{eqnarray}
(1+\alpha)\left(\frac{k_{\rm B}T_{\rm c}^{+}}{\hbar\bar{\omega}_{87}}\right)^{3}
[\gamma_{85}^{3}\zeta(3)
+\gamma_{\rm m}^{3}\mathcal{G}_{3}({\it z}_{\rm m})]-\alpha N_{\rm tot}=0, 
\end{eqnarray}
with 
${\it z}_{\rm m} = \exp[{-\delta/(k_{\rm B}T_{\rm c}^{+})}]$. 
Below $T_{\rm c}^{+}$, 
one has the following exact forms for two condensate populations, 
\begin{eqnarray}
\left \{
\begin{array}{lll}
N_{\rm c}^{87}
&=& 
\displaystyle{
\frac{1}{1+\alpha}N_{\rm tot}
}
- 
\displaystyle{
\left(\frac{k_{\rm B}T}{\hbar\bar{\omega}_{87}}\right)^{3}
}
[\zeta(3)+\gamma_{\rm m}^{3}\mathcal{G}_{3}({\it z}_{\rm m})],
\\
N_{\rm c}^{85}
&=& 
\displaystyle{
\frac{\alpha}{1+\alpha}N_{\rm tot} 
}
- 
\displaystyle{
\left(\frac{k_{\rm B}T}{\hbar\bar{\omega}_{87}}\right)^{3}
}
[\gamma_{85}^{3}\zeta(3)+\gamma_{\rm m}^{3}\mathcal{G}_{3}({\it z}_{\rm m})], 
\end{array}
\right.
\end{eqnarray}
where ${\it z}_{\rm m} = \exp{[-\delta/(k_{\rm B}T)]}$. 
We note that 
in the limit $T \rightarrow 0$, 
the above condensate populations 
correspond to the respective initial populations 
so that one has no heteronuclear molecules at $\delta > 0$ 
in the limit $T \rightarrow 0$. 
This is specific to the case of a Bose-Bose ideal gas mixture, 
where the ground state of the system 
only depends on the sign of the detuning. 
In contrast, 
the ground state of a gas including a fermionic component, 
which we will discuss in the following sections, 
is dominated by Fermi energy. 
One can have heteronuclear molecules even in the limit $T \rightarrow 0$ 
at positive detunings in such systems. 
Strictly speaking, 
these results in the limit $T \rightarrow 0$ are due to our treatment assuming 
$k_{\rm B}T \gg \hbar\bar{\omega}_{i}$. 
In the limit $T \rightarrow 0$ satisfying 
$k_{\rm B}T \lesssim \hbar\bar{\omega}_{i}$, 
the quantized discrete energy spectrum of each component cannot be ignored.

Secondly, we consider the case $\delta < 0$. 
According to the inequalities in Eq.~(\ref{eq17}) 
and $\delta < 0$, 
we find that 
$\mu_{2} $ cannot reach $0$. 
In other words, $^{85}$Rb atoms cannot be condensed ($N_{\rm c}^{85} = 0$), 
while BEC of heteronuclear molecules can take place 
under the condition $\delta = (1+\alpha)\mu_{2}$. 
This is reasonable because the lowest energy of heteronuclear molecules 
is lower than that of $^{85}$Rb atoms. 
One can determine $T_{\rm c}^{-}$ below which 
BEC of heteronuclear molecules takes place by solving the following equation,  
\begin{eqnarray}
(1+\alpha)
\left (\frac{k_{\rm B}T_{\rm c}^{-}}{\hbar\bar{\omega}_{87}}\right )^{3}
[\gamma_{\rm m}^{3}\zeta(3)
+\gamma_{85}^{3}\mathcal{G}_{3}({\it z}_{85})]-\alpha N_{\rm tot}=0, 
\end{eqnarray}
where 
${\it z}_{85} = \exp{[\delta/k_{\rm B}T_{\rm c}^{-}]}$. 
Below $T_{\rm c}^{-}$, we obtain the exact forms of two condensate populations, 
\begin{eqnarray}
\left \{
\begin{array}{lll}
N_{\rm c}^{87}
&=& 
\displaystyle{
\frac{1-\alpha}{1+\alpha}N_{\rm tot} 
}
- 
\displaystyle{
\left (\frac{k_{\rm B}T}{\hbar\bar{\omega}_{87}}\right )^{3}
}
[\zeta(3)+\gamma_{85}^{3}\mathcal{G}_{3}({\it z}_{85})], 
\\
N_{\rm c}^{\rm m}
&=& 
\displaystyle{
\frac{\alpha}{1+\alpha}N_{\rm tot} 
}
- 
\displaystyle{
\left (\frac{k_{\rm B}T}{\hbar\bar{\omega}_{87}}\right )^{3}
}
[\gamma_{\rm m}^{3}\zeta(3)+\gamma_{85}^{3}\mathcal{G}_{3}({\it z}_{85})], 
\end{array}
\right.
\end{eqnarray}
with 
${\it z}_{85} = \exp{[\delta/(k_{\rm B}T)]}$.

Finally, we consider the case $\delta=0$. 
Although 
BEC of $^{85}$Rb atoms 
and heteronuclear molecules can take place below  $T_{\rm c}^{87}$ 
when $\mu_{1}=\mu_{2}=\delta = 0$ is satisfied, 
this case has a problem 
in that condensate fractions cannot be uniquely determined 
because of the degeneracy of $^{87}$Rb atoms, $^{85}$Rb atoms and heteronuclear molecules. 
The transition temperature $T_{\rm c}^{0}$ below which 
BEC of $^{85}$Rb atoms and heteronuclear molecules appears, 
with $^{87}$Rb atoms being condensed, 
is given by 
\begin{eqnarray}
k_{\rm B}T_{\rm c}^{0} = 
\hbar\bar{\omega}_{87} 
\left [ \frac{\alpha}
{(\gamma_{85}^{3}+\gamma_{\rm m}^{3})
\zeta(3)(1+\alpha)}N_{\rm tot}\right ]^{1/3}. 
\end{eqnarray}
The condensate population $N_{\rm c}^{87}$ at $T = T_{\rm c}^{0}$ is given by 
\begin{eqnarray}
N_{\rm c}^{87} = N_{\rm tot}
-\left (\frac{k_{\rm B}T_{\rm c}^{0}}{\hbar\bar{\omega}_{87}}\right )^{3}\zeta(3)
\left (
1 +\gamma_{85}^{3} + 2\gamma_{\rm m}^{3}
\right ).
\end{eqnarray}

So far we have not specified trap frequency for each component. 
A trap frequency depends on a mass and a polarizability of a particle. 
The polarizability of a molecule is expressed by 
the sum of the atomic polarizabilities, 
when the molecular internuclear separation is 
large compared to the mean atomic radii~\cite{Morales}. 
Porlarizabilities of $^{87}$Rb atoms and $^{85}$Rb atoms 
are equal, because atoms are isotopic. 
In this isotopic case, 
ratios of trap frequencies $\gamma_{85}$ and 
$\gamma_{\rm m}$ are expressed by masses of two components 
according to simple formulae given in Ref.~\cite{Morales}. 
Two atomic masses are different, 
but the difference of masses are so small 
that the ratios of trap frequencies are almost unity; 
$\gamma_{85} \approx 0.99$ and $\gamma_{\rm m} \approx 0.99$. 
As a result, 
one is allowed to assume that 
all trap frequencies are equal 
$\bar{\omega}_{87} = \bar{\omega}_{85} = \bar{\omega}_{\rm m}$. 
We will use equal trap frequencies henceforth. 

\begin{figure}
\begin{center}
\includegraphics[width=8cm,height=8cm,keepaspectratio,clip]{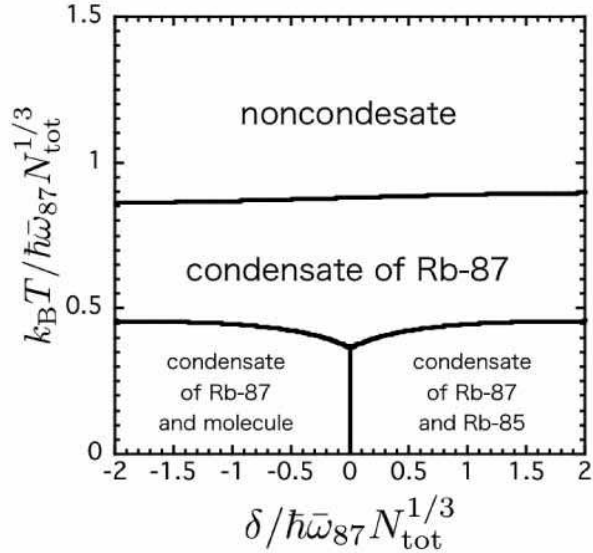}
\end{center}
\caption{
The phase diagram of $^{87}$Rb atoms, $^{85}$Rb atoms and heteronuclear molecules. 
We assume the ratio $\alpha = 2/15$, and equal trap frequencies 
$\bar{\omega}_{87} = \bar{\omega}_{85} = \bar{\omega}_{\rm m}$. 
}
\label{BBBTc.fig}
\end{figure}

\begin{figure}
\begin{center}
\includegraphics[width=16cm,height=19cm,keepaspectratio,clip]{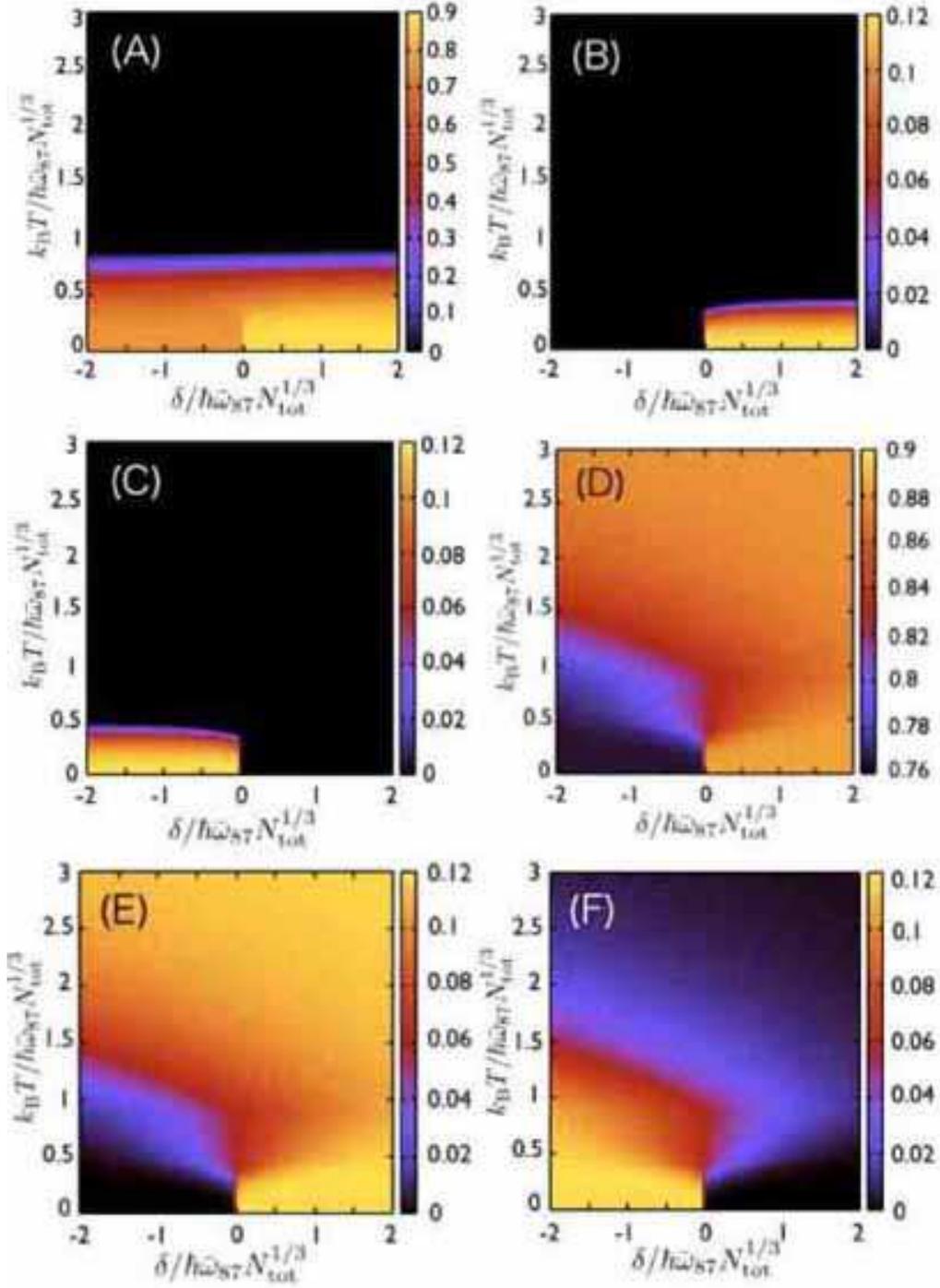}
\end{center}
\caption{
(Color online) 
(A) The condensate fraction of $^{87}$Rb atoms: $N_{\rm c}^{87}/N_{\rm tot}$. 
(B) The condensate fraction of $^{85}$Rb atoms: $N_{\rm c}^{85}/N_{\rm tot}$. 
(C) The condensate fraction of molecules: $N_{\rm c}^{\rm m}/N_{\rm tot}$. 
(D) The fraction of $^{87}$Rb atoms: $N_{87}/N_{\rm tot}$. 
(E) The fraction of $^{85}$Rb atoms: $N_{85}/N_{\rm tot}$. 
(F) The fraction of molecules: $N_{\rm m}/N_{\rm tot}$. 
We assume the ratio $\alpha = 2/15$, and equal trap frequencies 
$\bar{\omega}_{87} = \bar{\omega}_{85} = \bar{\omega}_{\rm m}$. 
}
\label{BBBN.fig}
\end{figure}

We plot the phase diagram for an ideal gas mixture of $^{87}$Rb atoms, 
$^{85}$Rb atoms and heteronuclear molecules in Fig.~\ref{BBBTc.fig} 
against the detuning $\delta$ and the temperature $T$. 
Fig.~\ref{BBBN.fig} shows population of each component. 
Figs.~\ref{BBBN.fig} (A),~\ref{BBBN.fig} (B) and~\ref{BBBN.fig} (C) show 
condensate fractions of $^{87}$Rb atoms, $^{85}$Rb atoms and 
heteronuclear Feshbach molecules, defined by 
$N_{\rm c}^{87}/N_{\rm tot}$, $N_{\rm c}^{85}/N_{\rm tot}$ and 
$N_{\rm c}^{\rm m}/N_{\rm tot}$. 
Figs.~\ref{BBBN.fig} (D),~\ref{BBBN.fig} (E) and~\ref{BBBN.fig} (F) show 
fractions of $^{87}$Rb atoms, $^{85}$Rb atoms and heteronuclear Feshbach molecules, 
defined by 
$N_{87}/N_{\rm tot}$, $N_{85}/N_{\rm tot}$ and 
$N_{\rm m}/N_{\rm tot}$. 
As discussed above, 
populations of $^{87}$Rb atoms, $^{85}$Rb atoms and 
heteronuclear molecules change dramatically and discontinuously 
on both sides of the $\delta=0$ at low temperatures. 

\begin{figure}
\begin{center}
\includegraphics[width=8cm,height=8cm,keepaspectratio,clip]{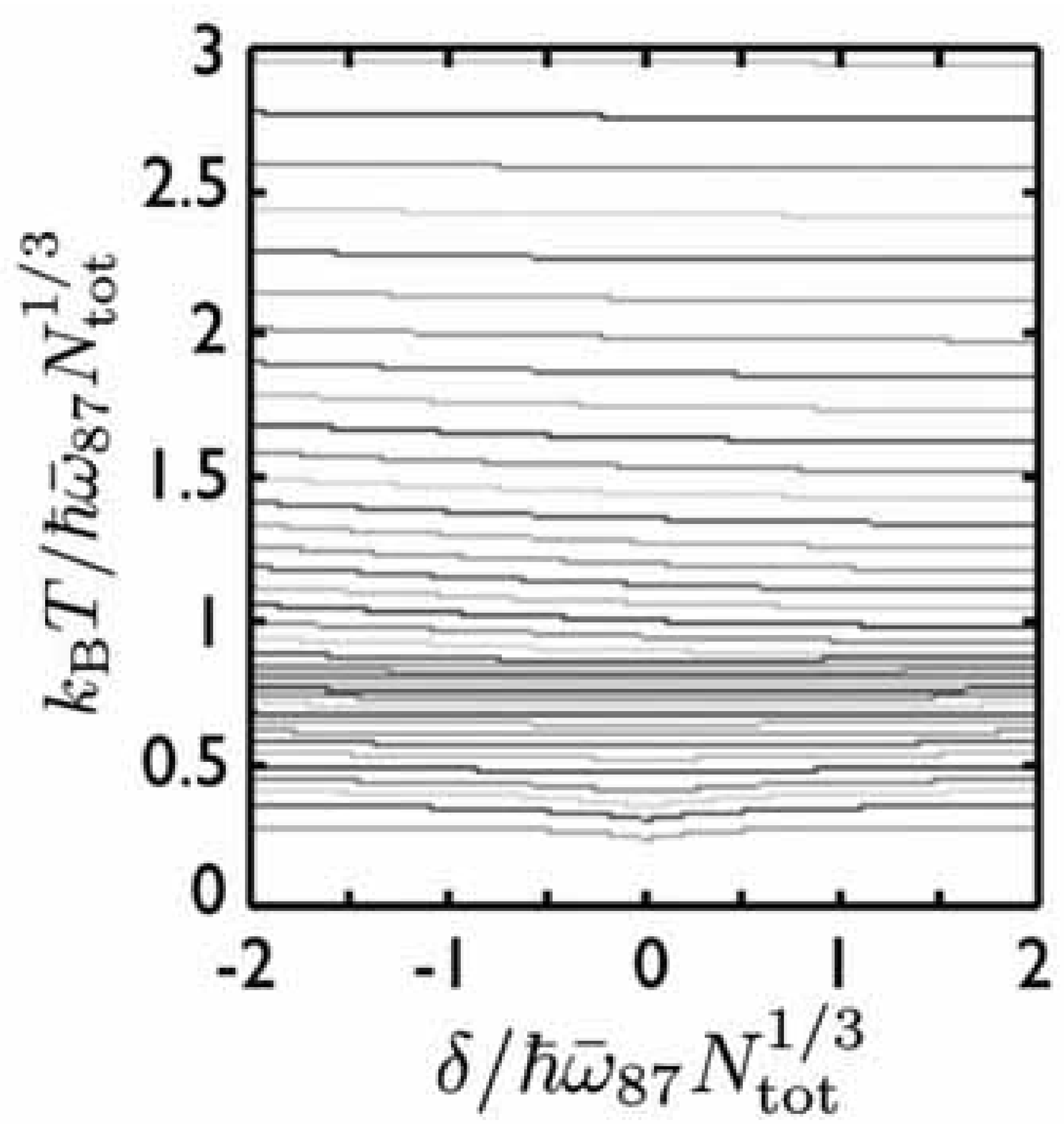}
\end{center}
\caption{
(Color online) 
The contours of constant entropy. 
We assume the ratio $\alpha = 2/15$, and equal trap frequencies 
$\bar{\omega}_{87} = \bar{\omega}_{85} = \bar{\omega}_{\rm m}$. 
}
\label{BBBS.fig}
\end{figure}

In order to form Feshbach molecules, 
an adiabatic ramp of the magnetic field is typically used. 
Throughout this process, the entropy should be conserved. 
The total entropy of this system is given by 
\begin{eqnarray}
S&=&
\tilde{S}_{\rm B}(z_{87})+\tilde{S}_{\rm B}(z_{85})
+\tilde{S}_{\rm B}(z_{\rm m}), 
\end{eqnarray} 
where the explicit formula of 
$\tilde{S}_{\rm B}$ is given in Eq.~(\ref{entropyBF}). 
In Fig.~\ref{BBBS.fig}, we plot 
the contours of constant entropy assuming 
the system traverses through the adiabatic ramp.

The adiabatic process connects 
an initial state with the final state by the equal entropy. 
Initial numbers of $^{87}$Rb atoms and $^{85}$Rb atoms are given by 
\begin{eqnarray}
N_{87,{\rm ini}} = \frac{1}{1+\alpha}N_{\rm tot},
&& 
N_{85,{\rm ini}} = \frac{\alpha}{1+\alpha}N_{\rm tot}. 
\end{eqnarray}
We define the initial transition temperatures for $^{87}$Rb atoms 
and $^{85}$Rb atoms by 
\begin{eqnarray}
k_{\rm B}T_{\rm c, ini}^{87}= 
\hbar\bar{\omega}_{87}
\left[\frac{N_{87,{\rm ini}}}{\zeta(3)}\right]^{1/3}, 
&&
k_{\rm B}T_{\rm c, ini}^{85}= 
\hbar\bar{\omega}_{85}
\left[\frac{N_{85,{\rm ini}}}{\zeta(3)}\right]^{1/3}. 
\end{eqnarray}
The transition temperatures satisfy 
$T_{\rm c, ini}^{87}>T_{\rm c, ini}^{85}$ 
since $^{87}$Rb is a majority component. 
For $T_{\rm c, {\rm ini}}^{85} \leq T_{{\rm ini}} \leq T_{\rm c, {\rm ini}}^{87}$, 
only BEC of $^{87}$Rb atoms occurs 
under the condition given by $\mu_{1} = \alpha\mu_{2}$. 
On the other hand, both $^{87}$Rb atoms and $^{85}$Rb atoms are condensed 
for $0 \leq T_{{\rm ini}} \leq T_{\rm c, {\rm ini}}^{85}$ under the condition 
$\mu_{1} = \mu_{2} = 0$. 
In this temperature region, 
the condensate populations are given by 
\begin{eqnarray}
\left \{
\begin{array}{lll}
N_{\rm c, {\rm ini}}^{87}&=&
\displaystyle{
\frac{1}{1+\alpha}N_{\rm tot}
}
-
\displaystyle{
\left(\frac{k_{\rm B}T_{{\rm ini}}}
{\hbar\bar{\omega}_{87}}\right)^{3}
}
\zeta(3), 
\\
N_{\rm c, {\rm ini}}^{85}&=&
\displaystyle{
\frac{\alpha}{1+\alpha}N_{\rm tot}
}
-
\displaystyle{
\left(\frac{k_{\rm B}T_{\rm ini}}{\hbar\bar{\omega}_{85}}\right)^{3}
}
\zeta(3). 
\end{array}
\right.
\end{eqnarray}

We now determine the conversion efficiency from the heteronuclear molecule fraction 
at zero detuning. 
As noted above, in the case of Bose-Bose ideal gas mixtures, 
one cannot uniquely determine the heteronuclear molecule fraction at $\delta = 0$ below $T_{\rm c}$. 
We thus adopt 
the molecular conversion efficiency 
as the heteronuclear molecule fraction 
just above zero detuning $\delta \rightarrow 0^{+}$; 
\begin{eqnarray}
\chi_{0} \equiv \frac{N_{\rm m}(\delta\rightarrow 0^{+})}
{N_{85, {\rm ini}}}
=\frac{N_{\rm m}(\delta\rightarrow 0^{+})}
{N_{\rm tot}}\frac{1+\alpha}{\alpha}.
\end{eqnarray}
This treatment is appropriate 
since the adiabatic sweep starts at $\delta>0$ side 
and a formation rate and a decay rate vanish 
at $\delta=0$~\cite{williams:2006}.

\begin{figure}
\begin{center}
\includegraphics[width=8cm,height=8cm,keepaspectratio,clip]{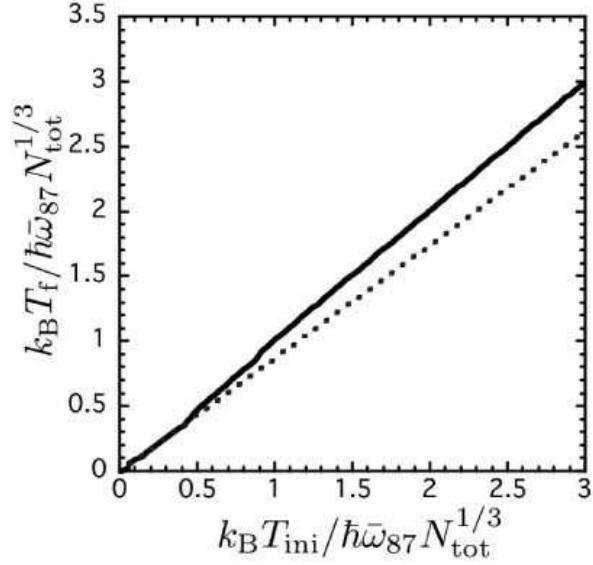}
\end{center}
\caption{
(Color online) 
The relation between the initial temperature ($\delta \rightarrow \infty$) 
and the final temperature ($\delta\rightarrow 0^{+}$). 
The solid line is the numerical result, 
while the dotted line 
is the result of the low temperature limit 
given in Eq.~(\ref{TfTiLowBBB}). 
We assume the ratio $\alpha = 2/15$ and equal trap frequencies 
$\bar{\omega}_{87} = \bar{\omega}_{85} = \bar{\omega}_{\rm m}$. 
}
\label{BBBTfTi.fig}
\end{figure}

By equating a final entropy $S_{\rm f}(T_{\rm f},\delta\rightarrow 0^{+})$ 
just above the zero detuning at $\delta \rightarrow 0^{+}$ 
with an initial entropy 
$S_{\rm ini}(T_{\rm ini})=\tilde{S}_{\rm B}(z_{87,{\rm ini}})
+\tilde{S}_{\rm B}(z_{85,{\rm ini}})$, 
we obtain the relation between 
the final temperature $T_{\rm f}$ and the initial temperature $T_{\rm ini}$. 
In the low temperature limit, 
we obtain the explicit formula, 
\begin{eqnarray}
T_{\rm f}=
\left (
\frac{1+\gamma_{85}^{3}}{1+\gamma_{85}^{3}+\gamma_{\rm m}^{3}} 
\right )^{1/3}
T_{\rm ini}. 
\label{TfTiLowBBB}
\end{eqnarray}
Fig.~\ref{BBBTfTi.fig} shows the relation between the initial temperature and 
the final temperature determined from 
$S_{\rm f}(T_{\rm f},\delta\rightarrow 0^{+}) = S_{\rm ini}(T_{\rm ini})$. 
The solid line is the numerical result and the dotted line 
is the low temperature limit given in Eq.~(\ref{TfTiLowBBB}). 
In the high temperature limit, 
the heteronuclear molecular population is so small 
that the contribution from the heteronuclear molecule entropy to the total entropy is small. 
As a result, one has $T_{\rm f} \approx T_{\rm ini}$ in the high temperature region. 

\begin{figure}
\begin{center}
\includegraphics[width=8cm,height=8cm,keepaspectratio,clip]{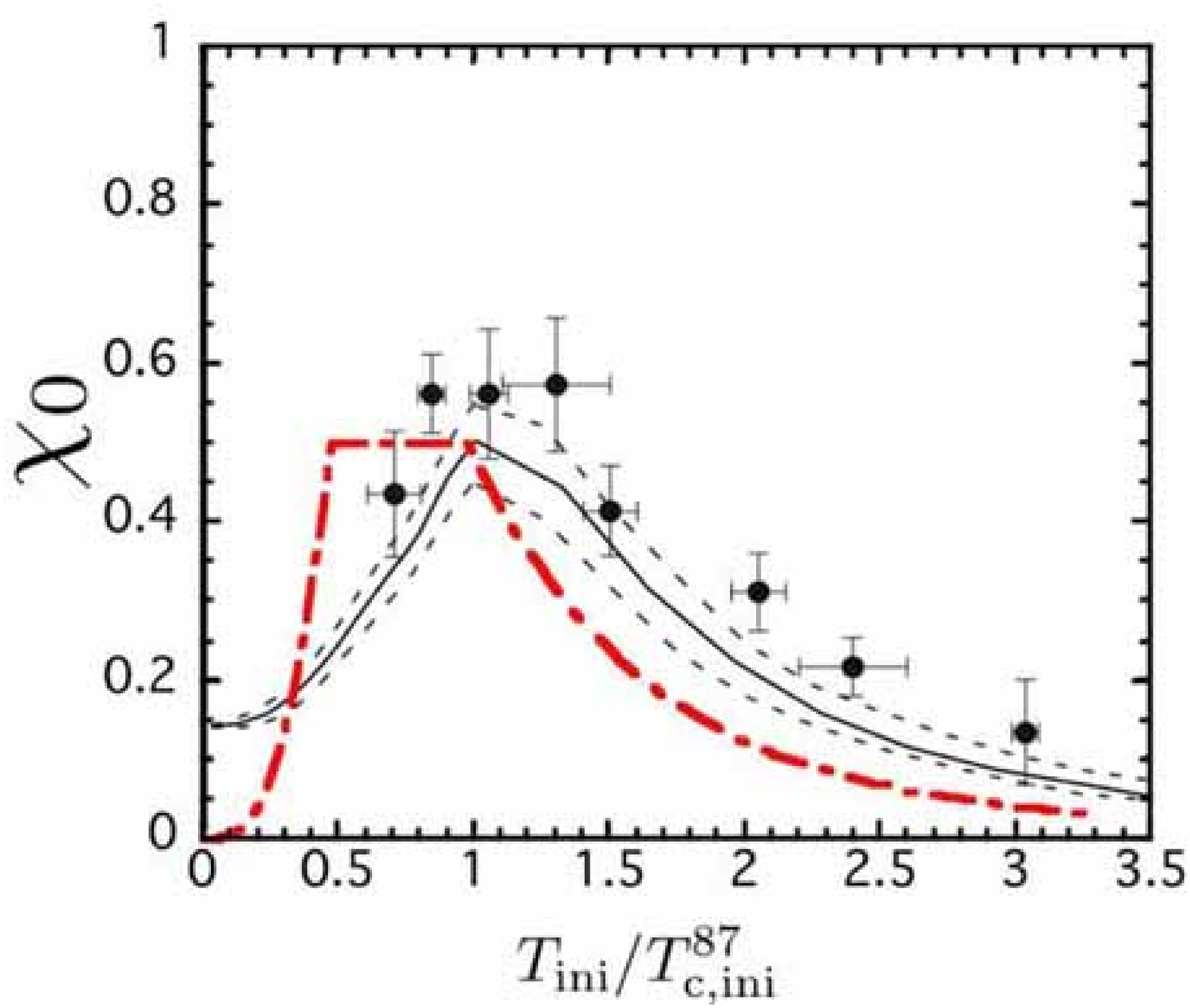}
\end{center}
\caption{
(Color online) 
The conversion efficiency as a function of the initial temperature. 
We assume the ratio $\alpha = 2/15$, and equal trap frequencies 
$\bar{\omega}_{87} = \bar{\omega}_{85} = \bar{\omega}_{\rm m}$. 
}
\label{BBBchi.fig}
\end{figure}

Fig.~\ref{BBBchi.fig} shows the molecular conversion efficiency 
as a function of the initial temperature. 
The dots are the experimental data~\cite{papp:2006}. 
The dot-dashed line is the result of our calculation. 
The solid line shows the result of the SPSS model 
and the dashed lines are its uncertainty~\cite{papp:2006}, 
where we note that a Maxwell-Boltzmann distribution is always used 
for $^{85}$Rb atoms in their simulation~\cite{papp:2006}. 
Our calculation agrees quite well with 
the behavior of the experiment~\cite{papp:2006} 
{\it without} any fitting parameters. 

In the experiment~\cite{papp:2006}, 
the ratio $\alpha$ is changed in order to vary $T/T_{\rm c, ini}^{87}$. 
However, the qualitative behavior of our calculation 
is unchanged even if we change the value $\alpha$. 
Especially, 
the conversion efficiency has a plateau at  
$\chi_{0} = \bar{\omega}_{85}^{3}/(\bar{\omega}_{85}^{3}+\bar{\omega}_{\rm m}^{3})$, 
where $^{87}$Rb atoms are condensed. 
This value does not depend on $\alpha$. 
The plateau is also seen in the experimental data. 

The origin of the deviation of our result from the experimental result 
may be due to our treatment using the ideal gas mixture. 
Including the mean-field interaction for the majority component 
$^{87}$Rb atoms in our model will shift the results. 

Below the plateau region, 
the conversion efficiency starts to decrease 
when both $^{85}$Rb and $^{87}$Rb atoms become Bose-condensed. 
Comparing our result with the experimental data, 
the data point at the lowest temperature ($\chi_{0} \simeq 40\%$) 
might be considered as the signal of  $^{85}$Rb BEC. 
We write down the conversion efficiency $\chi_{0}$ in the region 
where both $^{87}$Rb and $^{85}$Rb components are condensed, 
as a function of the initial temperature $T_{\rm ini}$, 
as follows 
\begin{eqnarray}
\chi_{0} = \left (\frac{k_{\rm B}T_{\rm f}}{\hbar\bar{\omega}_{\rm m}}\right)^{3}
\zeta(3)\frac{1}{N_{\rm tot}}\frac{1+\alpha}{\alpha}
= 
\frac{(1+\gamma_{85}^{3})\gamma_{\rm m}^{3}}
{1+\gamma_{85}^{3}+\gamma_{\rm m}^{3}}
\left (\frac{k_{\rm B}T_{\rm ini}}{\hbar\bar{\omega}_{87}}\right)^{3}
\zeta(3)\frac{1}{N_{\rm tot}}\frac{1+\alpha}{\alpha}. 
\label{BBB33eq}
\end{eqnarray}
For $T_{\rm ini} = 0$, one cannot convert any atoms 
into heteronuclear Feshbach molecules.

Although both conversion efficiencies in our model (dot-dashed line) 
and in the SPSS model~\cite{papp:2006} (solid line) 
drop at low temperatures as shown in Fig.~\ref{BBBchi.fig}, 
we note that mechanisms of the decrease in the efficiency are fundamentally different. 
In our model, 
molecules are not condensed in the region where the efficiency decreases, 
while $^{85}$Rb atoms are condensed since $\delta \rightarrow 0^{+}$. 
The total energy becomes lower when heteronuclear molecules decay into atoms 
and become Bose-condensed, 
because the lowest energy of $^{85}$Rb atoms is lower than that of 
heteronuclear molecules for $\delta \rightarrow 0^{+}$. 
This leads to the decrease of the conversion efficiency in our model. 
On the other hand, 
the decrease of the efficiency in the SPSS model 
is entirely due to the use of the Maxwell-Boltzmann distribution 
for a minority component of $^{85}$Rb atoms at all temperatures. 
In the region where BEC of $^{87}$Rb atoms takes place, 
$^{87}$Rb atoms are localized in the phase space, 
while $^{85}$Rb atoms spread in the phase space due to the classical gas distribution. 
Thus, in the SPSS model, it becomes difficult for $^{85}$Rb atoms to find partners with $^{87}$Rb atoms 
as temperature decreases, 
because of the principle of a monogamy clause for a proximity atomic pair in phase space. 
This leads to the decrease of the conversion efficiency in the SPSS model. 

On the other hand, 
when $^{85}$Rb gas is Bose condensed, 
the SPSS model will predict that the molecular conversion efficiency 
quickly approaches $100\%$ as the gases go to the ground state~\cite{papp:2006}. 
This is because 
every atom occupies the same state in phase space 
if the system is in the ground state, 
and $^{85}$Rb atoms can easily find partners with $^{87}$Rb atoms 
close to them in phase space 
due to the principle of the SPSS model. 
We note that our theory would have also 
predicted that the conversion efficiency reaches 100$\%$ 
if we defined the conversion efficiency as 
$\chi_{0}\equiv N_{\rm m}(\delta \rightarrow 0^{-})/N_{85,{\rm ini}}$ 
just below zero detuning. 
However, this treatment is inappropriate to the argument 
based on the energy and momentum conservation 
proposed by Williams {\it et al.}~\cite{williams:2006}.

The conversion efficiency $\chi_{0}$ in the plateau region, 
where only $^{87}$Rb atoms are condensed, 
can be determined as follows. 
We recall from Eq. (\ref{eq8}) 
that the fugacities satisfy $z_{87}z_{85}=z_{\rm m}$ at zero detuning. 
When $^{87}$Rb atoms are condensed, {\it i.e.}, $z_{87}=1$, 
one has the relation $z_{85}=z_{\rm m}$, 
which makes the population ratio of $^{85}$Rb atoms 
and heteronuclear molecules independent of the temperature. 
As a result, the conversion efficiency $\chi_{0}$ 
keeps the constant value, exhibiting the plateau. 
This maximum conversion does not depend on 
the atomic population ratio $\alpha$, 
but does depend on the trap frequencies for minority component and molecule; 
\begin{eqnarray}
\chi_{0,{\rm max}} = \frac{\bar{\omega}_{85}^{3}}
{\bar{\omega}_{85}^{3}+\bar{\omega}_{\rm m}^{3}}. 
\label{31eq}
\end{eqnarray}  
Even if we regard $^{85}$Rb atoms and heteronuclear molecules 
as classical gases, 
this mechanism does not change and the conversion efficiency has a plateau 
when $^{87}$Rb atoms are condensed, {\it i.e.}, $z_{87}=1$. 
This result is also different from the result of the SPSS model, 
which shows that the efficiency decreases in the low temperature region 
when $^{85}$Rb atoms are regarded as the classical gas. 
From Eq. (\ref{31eq}), 
one sees that the maximum conversion efficiency in our model 
is given by $\chi_{0,{\rm max}} = 50\%$, 
for equal trap frequencies 
$\bar{\omega}_{87} = \bar{\omega}_{85} = \bar{\omega}_{\rm m}$.

Before closing this section, 
we comment on the conversion efficiencies of bosonic {\it homonuclear} Feshbach molecules 
composed of bosonic atoms. 
In this case, 
the theoretical result 
using the classical gas approximation based on the theoretical model 
by Williams {\it et al}.~\cite{williams:2006} 
is qualitatively consistent 
with the result of experiment 
and the SPSS model~\cite{hodby:2005}. 
However, when we extend Ref.~\cite{williams:2006} to the low temperature regime 
where Bose-Einsten condensation appears, 
the result is expected to be also completely different 
from the result by the SPSS model. 
In the SPSS model, 
one expects a complete conversion 
as long as Bose-Einstein condensation of atomic component appears. 
On the other hand, 
applying the theoretical model by Williams {\it et al}.~\cite{williams:2006}, 
one finds that the conversion efficiency 
does not have a plateau 
but starts to decrease when BEC of atoms appears. 
As a result, one cannot convert any atoms into molecules in the limit $T_{\rm ini} \rightarrow 0$. 
This result can be obtained from 
the fugacity relation at zero detuning $z_{\rm a}^{2} = z_{\rm m}$, 
where $z_{\rm a}$ is the fugacity of atomic component, 
and the definition of the molecular conversion efficiency 
using the molecular population just above zero detuning $N_{\rm m}(\delta \rightarrow 0^{+})$. 

\section{Gas of Majority Bosonic Atom, Minority Fermionic Atom 
and its Heteronuclear Molecule  \label{SecBFF}}

In this section, we discuss the conversion efficiency 
of heteronuclear Feshbach molecules 
composed of bosonic and fermionic atoms. 
We first note that 
the phase diagram of an ideal Bose-Fermi mixture gas 
with a majority bosonic component is not smoothly connected 
to the case of a majority fermionic component. 
When a majority atomic component is bosonic, 
not all bosonic atoms are converted into heteronuclear molecules 
and thus the remaining atoms can undergo 
Bose-Einstein condensation at any detuning. 
In contrast, 
in the case of a majority fermionic component, 
the transition temperature of Bose-Einstein condensation drops to zero 
at a certain detuning where all bosonic atoms are converted into heteronuclear molecules. 
The phase diagram of a population balanced gas of 
Bose-Fermi mixture belongs to the latter case. 
Morales {\it et al.} studied 
populations and adiabatic trajectories in $\delta$-$T$ plane 
for an ideal trapped gas of bosonic and fermionic atoms and 
the heteronuclear molecules formed from these atomic components~\cite{Morales}. 
Their numerical results assumed equal populations in the two atomic components 
and different trap frequencies. 

In this section, we discuss the case 
with a majority bosonic atom: 
$\{{\rm B}_{>}+{\rm F}_{<}\leftrightarrow {\rm F}_{\rm m}\}$, 
following the same procedure as in the previous section. 
As noted above, 
the calculations given in this section is only valid for $\alpha < 1$. 
On the other hand, 
the population balanced gas of Bose-Fermi mixture characterized by $\alpha = 1$ 
is smoothly connected to the case with a majority fermionic component: 
$\{{\rm F}_{>}+{\rm B}_{<}\leftrightarrow {\rm F}_{\rm m}\}$, 
which we will discuss in the next section. 
 
\begin{figure}
\begin{center}
\includegraphics[width=8cm,height=8cm,keepaspectratio,clip]{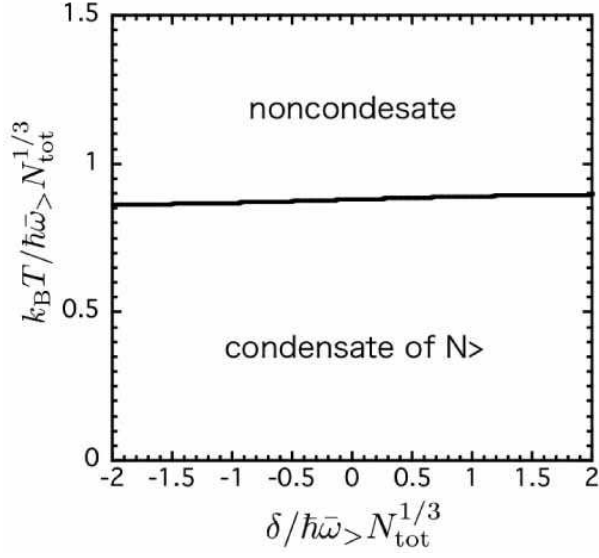}
\end{center}
\caption{
The phase diagram of the transition temperature of a bosonic majority component. 
We assume the ratio $\alpha = 2/15$, and equal trap frequencies 
$\bar{\omega}_{>} = \bar{\omega}_{<} = \bar{\omega}_{\rm m}$. 
}
\label{BFFTc.fig}
\end{figure}

\begin{figure}
\begin{center}
\includegraphics[width=15cm,height=15cm,keepaspectratio,clip]{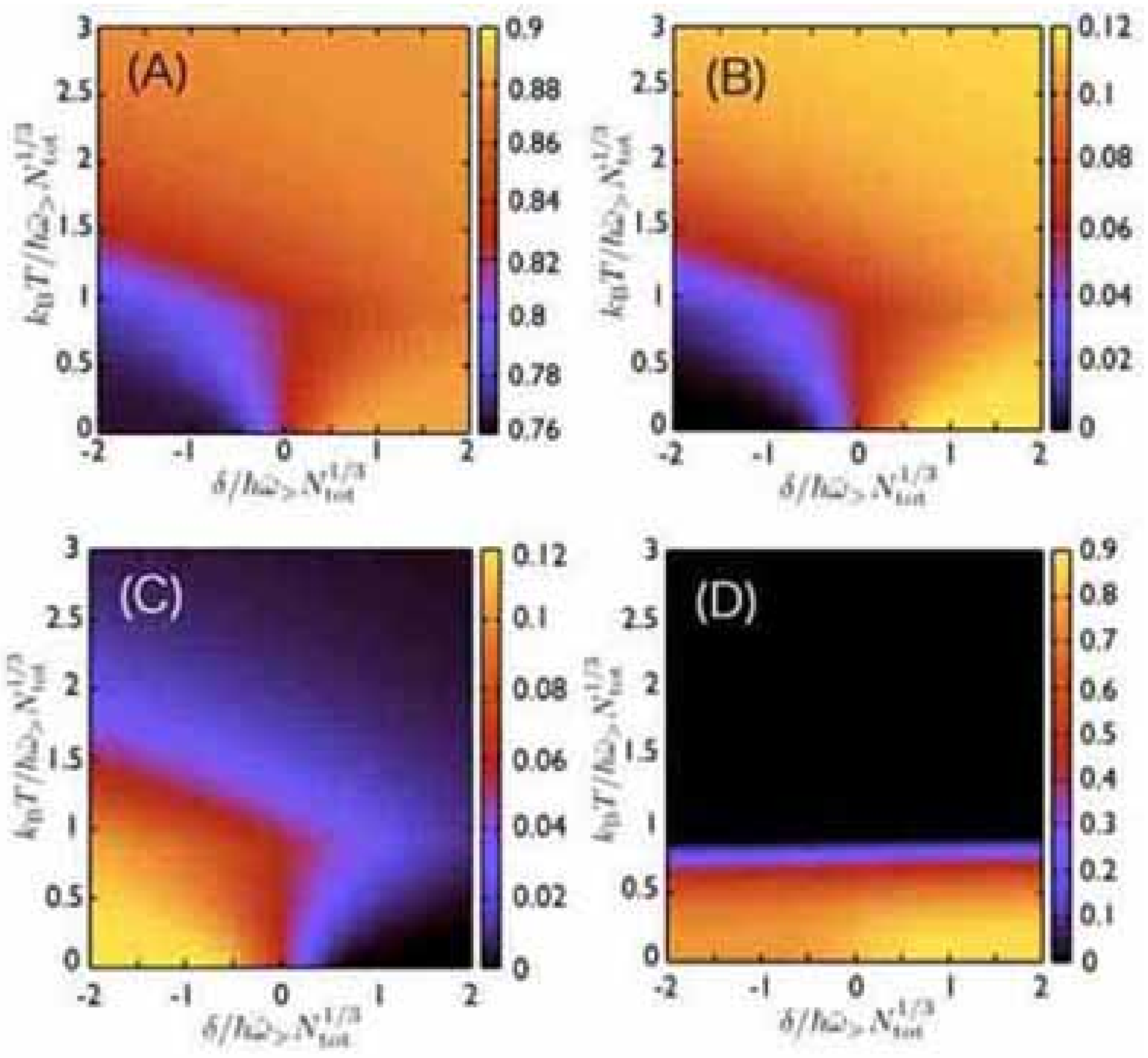}
\end{center}
\caption{
(Color online) 
(A) The fraction of a bosonic majority component: $N_{>}/N_{\rm tot}$. 
(B) The fraction of a fermionic minority component: $N_{<}/N_{\rm tot}$. 
(C) The fraction of a fermionic heteronuclear molecular component: $N_{\rm m}/N_{\rm tot}$. 
(D) The condensate fraction of a bosonic majority component: $N_{\rm c}^{>}/N_{\rm tot}$. 
We assume the ratio $\alpha = 2/15$, and equal trap frequencies 
$\bar{\omega}_{>} = \bar{\omega}_{<} = \bar{\omega}_{\rm m}$.
}
\label{BFFN.fig}
\end{figure}

\begin{figure}
\begin{center}
\includegraphics[width=8cm,height=8cm,keepaspectratio,clip]{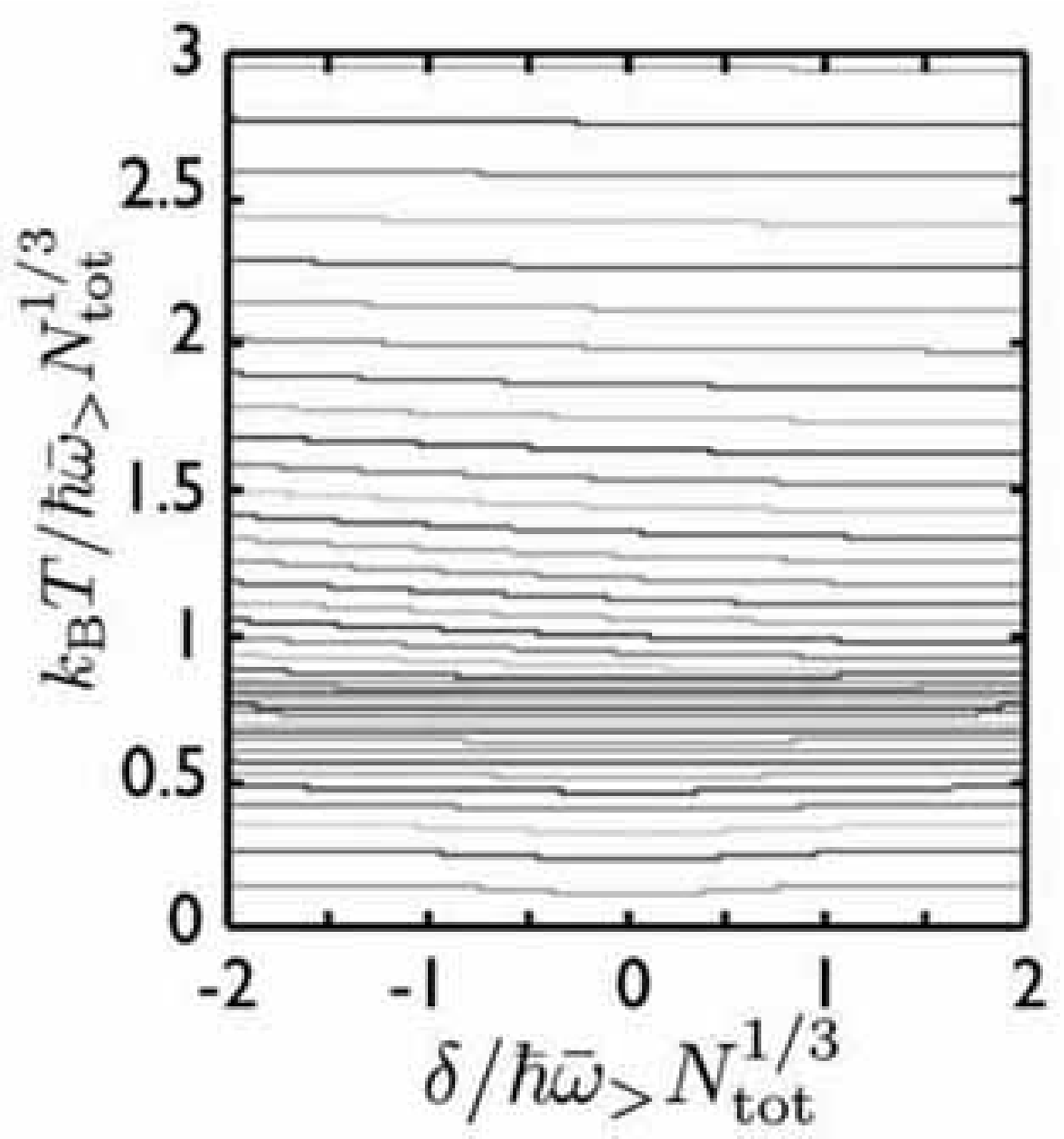}
\end{center}
\caption{
(Color online) 
The contours of constant entropy. 
We assume the ratio $\alpha = 2/15$, and equal trap frequencies 
$\bar{\omega}_{>} = \bar{\omega}_{<} = \bar{\omega}_{\rm m}$. 
}
\label{BFFS.fig}
\end{figure}

Under the two constraints in Eqs. (\ref{eq2}) and (\ref{eq4}), 
the populations of three components are given by 
\begin{eqnarray}
N_{>}=
N_{\rm c}^{>}+
\displaystyle{ 
\left(\frac{k_{\rm B}T}{\hbar\bar{\omega}_{>}}\right)^{3}
}
\mathcal{G}_{3}(z_{>}),&
N_{<}=
\displaystyle{
\left(\frac{k_{\rm B}T}{\hbar\bar{\omega}_{<}}\right)^{3}
}
\mathcal{F}_{3}(z_{<}),&
N_{\rm m}=
\displaystyle{
\left(\frac{k_{\rm B}T}{\hbar\bar{\omega}_{\rm m}}\right)^{3}
}
\mathcal{F}_{3}(z_{\rm m}),  
\end{eqnarray}
where
$N_{\rm c}^{>}$ is the number of condensate atoms in a bosonic majority component. 
The condition for a majority component BEC is $\mu_{1}=\alpha\mu_{2}$. 
Fig.~\ref{BFFTc.fig} shows 
the phase diagram showing the boundary of condensate phase of 
a bosonic majority component. 
Fig.~\ref{BFFN.fig} shows fraction of each component, where we assume $\alpha = 2/15$. 
We also assume equal trap frequencies 
$\bar{\omega}_{>} = \bar{\omega}_{<} = \bar{\omega}_{\rm m}$ for simplicity. 
Figs.~\ref{BFFN.fig} (A),~\ref{BFFN.fig} (B) and~\ref{BFFN.fig} (C) show 
fractions of the majority atomic, the minority atomic and the heteronuclear molecular 
component, defined by $N_{>}/N_{\rm tot}$, $N_{<}/N_{\rm tot}$ and 
$N_{\rm m}/N_{\rm tot}$. 
Fig.~\ref{BFFN.fig} (D) shows 
the condensed fraction of the majority atomic component defined by 
$N_{\rm c}^{>}/N_{\rm tot}$. 
Fig.~\ref{BFFS.fig} shows the contours of constant entropy 
assuming the system traverses through the adiabatic ramp, 
where the total entropy is given by 
$S = \tilde{S}_{\rm B}(z_{>})+S_{\rm F}(z_{<})+S_{\rm F}(z_{\rm m})$.

Under the condition $\mu_{1}=\alpha\mu_{2}$, 
the BEC transition temperature $T_{\rm c}$ 
at zero detuning $\delta = 0$ is given by 
\begin{eqnarray}
k_{\rm B}T_{\rm c}&=&\hbar\bar{\omega}_{>}
\left[\frac{\gamma_{<}^{3}+(1-\alpha)\gamma_{\rm m}^{3}}
{\gamma_{<}^{3}+\gamma_{\rm m}^{3}}
\frac{1}{(1+\alpha)\zeta(3)} N_{\rm tot} \right]^{1/3}.
\end{eqnarray}
Right at zero detuning $\delta=0$ and below $T_{\rm c}$, 
the populations of the bosonic majority component BEC $N_{\rm c}^{>}$, the fermionic heteronuclear molecule $N_{\rm m}$, 
and the fermionic minority component $N_{<}$ are given by 
\begin{eqnarray}
\left \{
\begin{array}{lll}
N_{\rm c}^{>}(\delta=0)
&=&
\displaystyle{
\frac{1}{1+\alpha} 
\frac{\gamma_{<}^{3}+(1-\alpha)\gamma_{\rm m}^{3}}
{\gamma_{<}^{3} + \gamma_{\rm m}^{3}} N_{\rm tot}
}
-
\displaystyle{
\left(\frac{k_{\rm B}T}{\hbar\bar{\omega}_{>}}\right)^{3}\zeta(3)
}, 
\\
N_{\rm m}(\delta=0)
&=&
\displaystyle{
\frac{\alpha}{1+\alpha}
}
\displaystyle{
\frac{\gamma_{\rm m}^{3}}{\gamma_{<}^{3}+\gamma_{\rm m}^{3}}
N_{\rm tot}
}, 
\\
N_{<}(\delta=0)
&=&N_{<,{\rm ini}}-N_{\rm m}(\delta=0)=
\displaystyle{
\frac{\alpha}{1+\alpha}
}
\displaystyle{
\frac{\gamma_{\rm <}^{3}}{\gamma_{<}^{3}+\gamma_{\rm m}^{3}}
N_{\rm tot}
}.
\end{array}
\right.
\end{eqnarray}

In the low temperature limit, 
one can obtain an analytical expression 
for the final entropy at zero detuning $\delta=0$ 
as  
\begin{eqnarray}
S_{\rm f} &= &\tilde{S}_{\rm B}(z_{>,{\rm f}}) + S_{\rm F}(z_{<,{\rm f}}) + S_{\rm F}(z_{\rm m,f})
\nonumber
\\
&\approx& 4k_{\rm B}\left(\frac{k_{\rm B}T_{\rm f}}
{\hbar\bar{\omega}_{>}}\right)^{3}\zeta(4)
+ k_{\rm B}\pi^{2}\left[\frac{(N_{<})^{2}}{6}\right]^{1/3}
\frac{k_{\rm B}T_{\rm f}}{\hbar\bar{\omega}_{<}}
+ k_{\rm B}\pi^{2}\left[\frac{(N_{\rm m})^{2}}{6}\right]^{1/3}
\frac{k_{\rm B}T_{\rm f}}{\hbar\bar{\omega}_{\rm m}}
\nonumber
\\
&\approx&
k_{\rm B}\pi^{2} (\gamma_{<}^{3}+\gamma_{\rm m}^{3}) 
\left[\frac{1}{6}
\left(\frac{\alpha}{1+\alpha}\frac{1}{\gamma_{<}^{3}+\gamma_{\rm m}^{3}}
N_{\rm tot}\right)^{2}\right]^{1/3}
\frac{k_{\rm B}T_{\rm f}}{\hbar\bar{\omega}_{>}},  
\label{35eq}
\end{eqnarray}
where the approximation leading to the last line of Eq. (\ref{35eq}) 
is valid when the final temperature satisfies the condition 
\begin{eqnarray}
k_{\rm B}T_{\rm f} \ll
\sqrt{\frac{1}{4\zeta (4)}}\pi
\left ( \frac{\gamma_{<}^{3}+\gamma_{\rm m}^{3}}{6} \right )^{1/6}
\left ( \frac{\alpha}{1+\alpha} N_{\rm tot}\right )^{1/3}
\hbar\bar{\omega}_{>}. 
\end{eqnarray}

In a pure atomic gas before a sweep of the magnetic field, 
initial populations in normal phase are given by 
\begin{eqnarray}
\left \{
\begin{array}{lllll}
N_{>,{\rm ini}}&=&
\displaystyle{ 
\frac{1}{1+\alpha}
}
N_{\rm tot}&=&
\displaystyle{ 
\left(\frac{k_{\rm B}T_{\rm ini}}{\hbar\bar{\omega}_{>}}\right)^{3}
}
\mathcal{G}_{3}(z_{>,{\rm ini}}),
\\
\\
N_{<,{\rm ini}}&=&
\displaystyle{ 
\frac{\alpha}{1+\alpha}
}
N_{\rm tot}&=&
\displaystyle{
\left(\frac{k_{\rm B}T_{\rm ini}}{\hbar\bar{\omega}_{<}}\right)^{3}
}
\mathcal{F}_{3}(z_{<,{\rm ini}}). 
\end{array}
\right.
\end{eqnarray}
The transition temperature $T_{\rm c, ini}$ and 
a condensate population $N_{\rm c, ini}^{>}$ below $T_{\rm c, ini}$ are given by  
\begin{eqnarray}
\left \{
\begin{array}{lll}
k_{\rm B}T_{\rm c, ini}&=&\hbar\bar{\omega}_{>}
\displaystyle{
\left(\frac{N_{>,{\rm ini}}}{\zeta(3)}\right)^{1/3}
},
\\
N_{\rm c, ini}^{>}&=&
\displaystyle { 
\frac{1}{1+\alpha}N_{\rm tot} 
}
- 
\displaystyle { 
\left(\frac{k_{\rm B}T_{\rm ini}}{\hbar\bar{\omega}_{>}}\right)^{3}\zeta(3)
}. 
\end{array}
\right.
\end{eqnarray}
In the low temperature limit, the initial entropy is given by 
\begin{eqnarray}
S_{\rm ini} &= &\tilde{S}_{\rm B}(z_{>,{\rm ini}}) + S_{\rm F}(z_{<,{\rm ini}})
\nonumber
\\
&\approx& 4k_{\rm B}\left(\frac{k_{\rm B}T_{\rm ini}}
{\hbar\bar{\omega}_{>}}\right)^{3}\zeta(4)
+ k_{\rm B}\pi^{2}\left[\frac{(N_{<,{\rm ini}})^{2}}{6}\right]^{1/3}
\frac{k_{\rm B}T_{\rm ini}}{\hbar\bar{\omega}_{<}}
\nonumber
\\
&\approx&
k_{\rm B}\pi^{2}\left[\frac{1}{6}
\left(\frac{\alpha}{1+\alpha}N_{\rm tot}\right)^{2}\right]^{1/3}
\gamma_{<}
\frac{k_{\rm B}T_{\rm ini}}{\hbar\bar{\omega}_{>}},
\label{39eq}
\end{eqnarray}
where the approximation leading to the last line of Eq. (\ref{39eq}) is valid when the initial temperature 
satisfies the condition 
\begin{eqnarray}
k_{\rm B}T_{\rm ini} \ll
\sqrt{\frac{1}{4\zeta (4)}}\pi
\left ( \frac{1}{6} \right )^{1/6}
\left ( \frac{\alpha}{1+\alpha} N_{\rm tot}\right )^{1/3}
\gamma_{<}^{1/2} 
\hbar\bar{\omega}_{>}. 
\end{eqnarray}

\begin{figure}
\begin{center}
\includegraphics[width=8cm,height=8cm,keepaspectratio,clip]{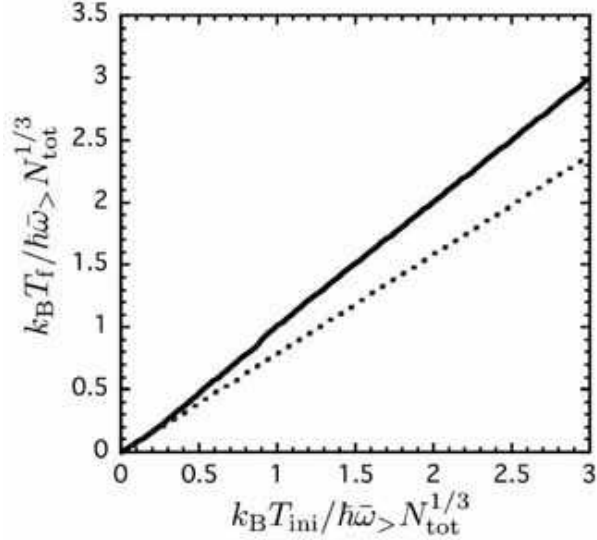}
\end{center}
\caption{
(Color online) 
The relation between the initial temperature and the final temperature at $\delta = 0$. 
The solid line is the numerical result 
and the dotted line is the analytical result in the low temperature limit given in Eq. (\ref{TfTiLowBFF}). 
We assume the ratio $\alpha = 2/15$, and equal trap frequencies 
$\bar{\omega}_{>} = \bar{\omega}_{<} = \bar{\omega}_{\rm m}$. 
}
\label{BFFTfTi.fig}
\end{figure}

Connecting the final entropy at zero detuning $\delta = 0$ with the initial entropy, 
we obtain the relation between an initial temperature $T_{\rm ini}$ and 
a final temperature $T_{\rm f}$. 
Especially in the low temperature limit, 
using Eqs. (\ref{35eq}) and (\ref{39eq}), 
we obtain the explicit relation 
\begin{eqnarray}
T_{\rm f}=
\frac{\gamma_{<}}{(\gamma_{<}^{3}+\gamma_{\rm m}^{3})^{1/3}}T_{\rm ini}. 
\label{TfTiLowBFF}
\end{eqnarray}
In the high temperature limit, 
the heteronuclear molecular population is so small 
that the contribution of the heteronuclear molecule entropy to the total entropy is very small. 
As a result, one has $T_{\rm f} \approx T_{\rm ini}$ in the high temperature region. 
Fig.~\ref{BFFTfTi.fig} shows the numerical result 
for the final temperature as a function of the initial temperature. 
For comparison, we also plot the analytical result in 
the low temperature limit given in Eq. (\ref{TfTiLowBFF}).

\begin{figure}
\begin{center}
\includegraphics[width=8cm,height=8cm,keepaspectratio,clip]{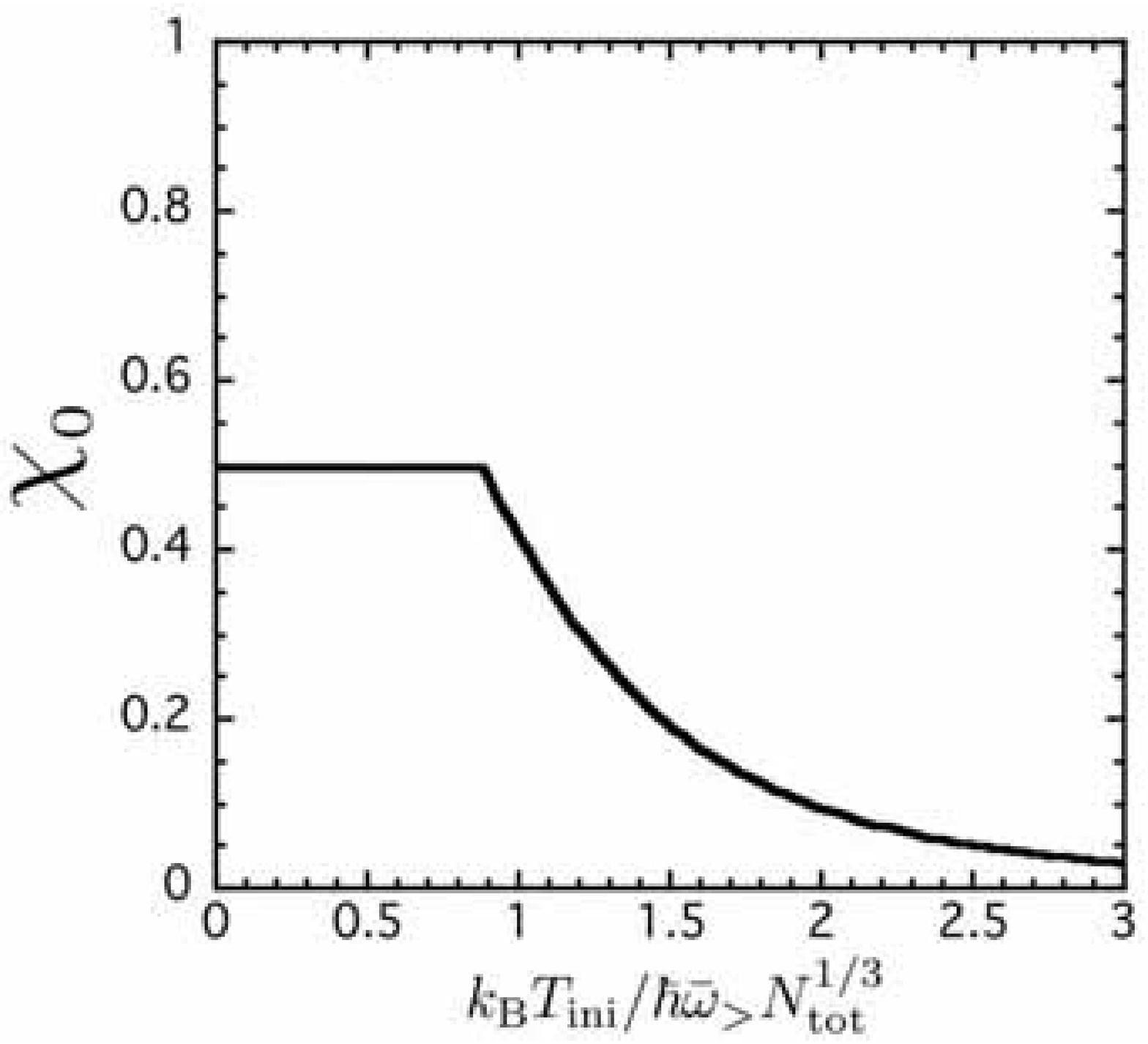}
\end{center}
\caption{
The molecular conversion efficiency as a function of the initial temperature. 
We assume the ratio $\alpha = 2/15$, and equal trap frequencies 
$\bar{\omega}_{>} = \bar{\omega}_{<} = \bar{\omega}_{\rm m}$. 
}
\label{BFFchi.fig}
\end{figure}

Fig.~\ref{BFFchi.fig} shows the conversion efficiency as a function of the initial temperature, 
calculated from the molecular population at zero detuning $\delta = 0$. 
The molecular conversion efficiency $\chi_{0}$ in a plateau region, 
where the bosonic majority component is Bose condensed, is given by 
\begin{eqnarray}
\chi_{0} = \frac{N_{\rm m}(\delta=0)}{N_{\rm tot}}
\frac{1+\alpha}{\alpha}
=
\frac{\bar{\omega}_{<}^{3}}{\bar{\omega}_{<}^{3}+\bar{\omega}_{\rm m}^{3}}. 
\end{eqnarray}
As shown in Fig.~\ref{BFFchi.fig}, 
this value is the maximum conversion efficiency, 
which does not depend on the number ratio $\alpha$. 
From the relation between fugacities $z_{>}z_{<}=z_{\rm m}$ 
at zero detuning $\delta = 0$, 
we find $z_{<}=z_{\rm m}$ when the majority component is Bose condensed, 
{\it i.e.}, $z_{>}=1$. 
This makes number ratio of the minority component and 
the heteronuclear molecule constant, 
and thus the molecular conversion efficiency exhibits a plateau. 

We note that this result is not expected 
from the analysis using the SPSS model. 
As the system approaches the ground state, 
bosonic atoms occupy the same state in the phase space. 
In contrast, fermionic atoms spread forming the Fermi surface 
due to Pauli exclusion principle. 
At low temperatures where the quantum degeneracy appears, 
the SPSS model predicts that 
the conversion efficiency decreases 
as the temperature goes to zero, 
because the number of atoms that can find a pair 
decreases 
due to a monogamy clause for a proximity atomic pair 
in phase space. 
The prescription disposing the atoms of different species 
in the same phase space leads to this result.

\section{Gas of Majority Fermionic Atom, Minority Bosonic Atom 
and its Heteronuclear Molecule  \label{SecFBF}}

\begin{figure}
\begin{center}
\includegraphics[width=8cm,height=8cm,keepaspectratio,clip]{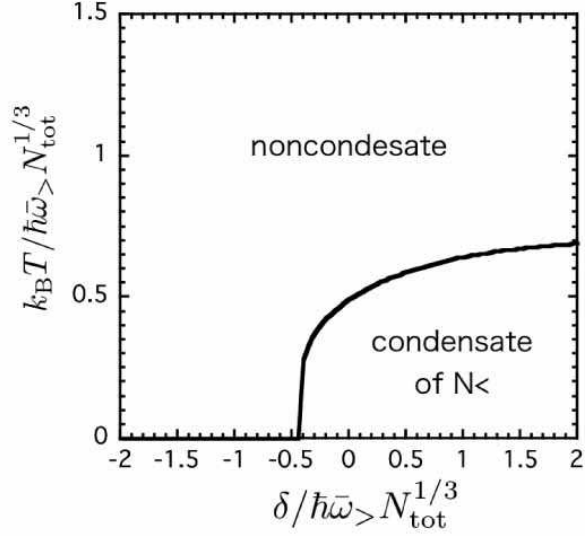}
\end{center}
\caption{
The phase diagram of the transition temperature of a bosonic minority  component. 
We assume the ratio $\alpha = 3/4$, and equal trap frequencies 
$\bar{\omega}_{>} = \bar{\omega}_{<} = \bar{\omega}_{\rm m}$. 
}
\label{FBFTc.fig}
\end{figure}

\begin{figure}
\begin{center}
\includegraphics[width=15cm,height=15cm,keepaspectratio,clip]{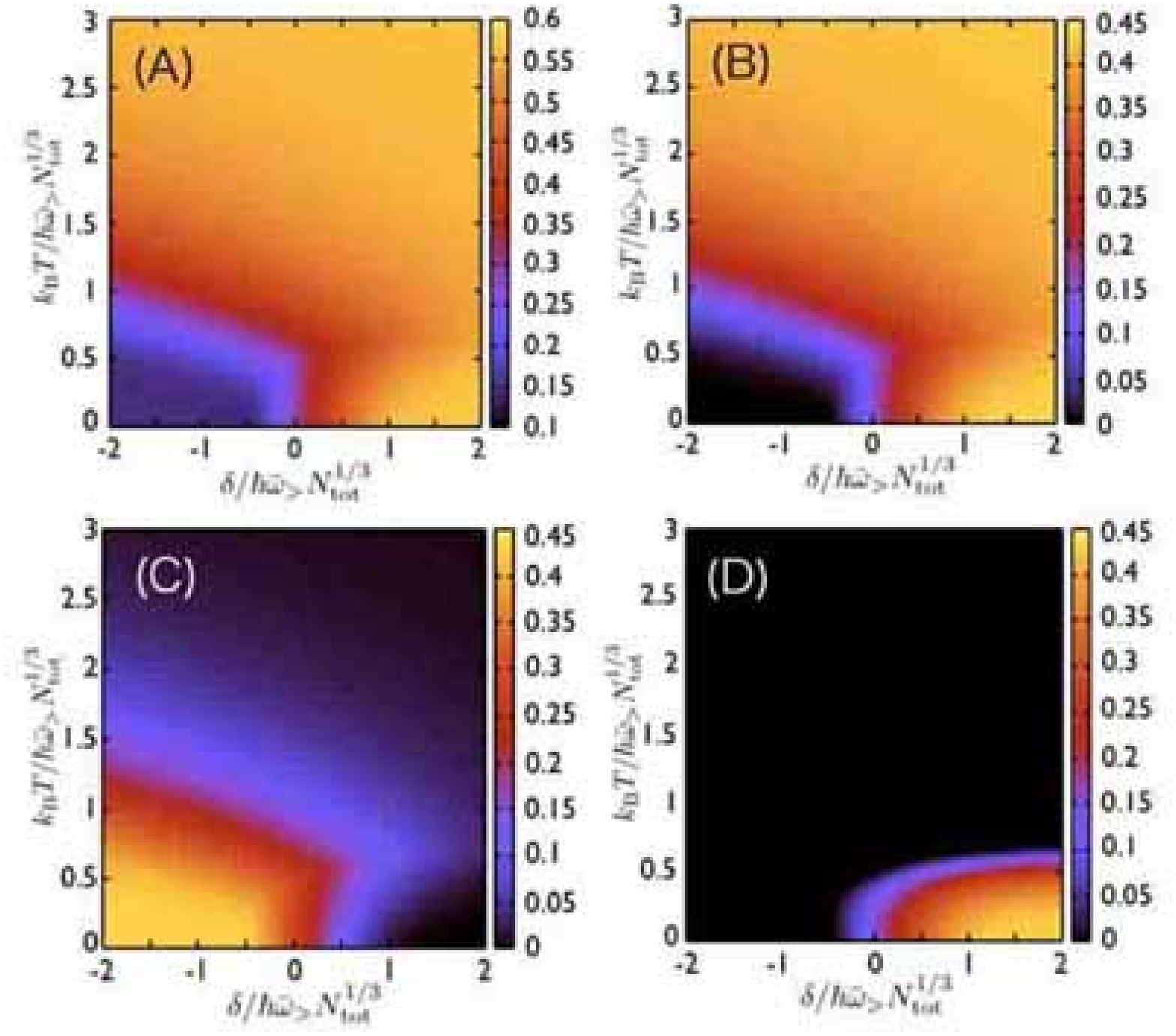}
\end{center}
\caption{
(Color online) 
(A) The fraction of a fermionic majority component: $N_{>}/N_{\rm tot}$. 
(B) The fraction of a bosonic minority component: $N_{<}/N_{\rm tot}$. 
(C) The fraction of a fermionic heteronuclear molecular component: $N_{\rm m}/N_{\rm tot}$. 
(D) The condensate fraction of a bosonic minority component: $N_{\rm c}^{<}/N_{\rm tot}$. 
We assume the ratio $\alpha = 3/4$, and equal trap frequencies 
$\bar{\omega}_{>} = \bar{\omega}_{<} = \bar{\omega}_{\rm m}$. 
}
\label{FBFN.fig}
\end{figure}

\begin{figure}
\begin{center}
\includegraphics[width=8cm,height=8cm,keepaspectratio,clip]{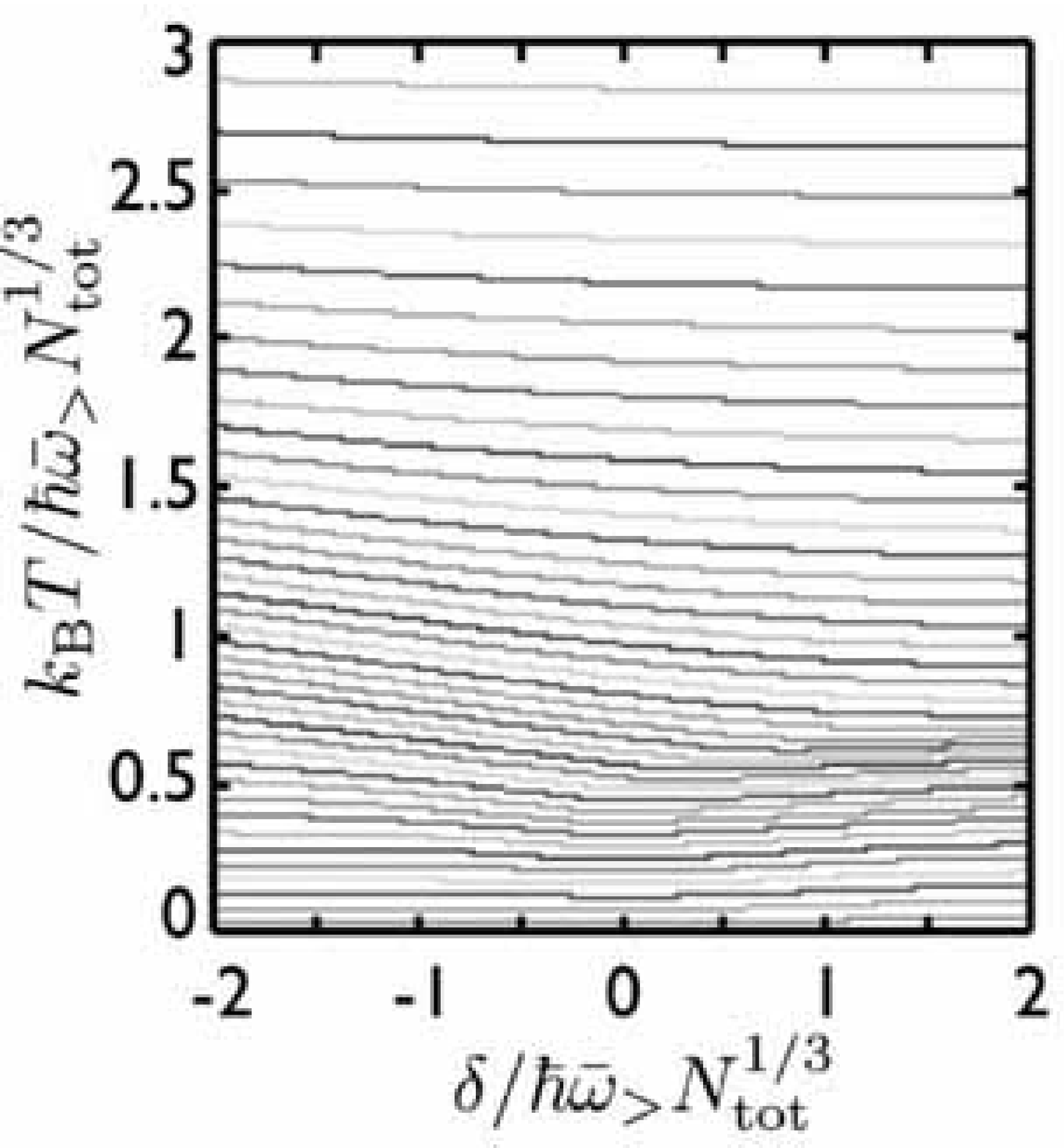}
\end{center}
\caption{
(Color online) 
The contours of constant entropy. 
We assume the ratio $\alpha = 3/4$, and equal trap frequencies 
$\bar{\omega}_{>} = \bar{\omega}_{<} = \bar{\omega}_{\rm m}$. 
}
\label{FBFS.fig}
\end{figure}

In this section, we consider the third case, 
with a majority component of fermionic atoms: $\{{\rm F}_{>}+{\rm B}_{<}\leftrightarrow {\rm F}_{\rm m}\}$. 
Under the two constraints in Eqs. (\ref{eq2}) and (\ref{eq4}), 
the populations of three components are given by 
\begin{eqnarray}
N_{>}=
\left(\frac{k_{\rm B}T}{\hbar\bar{\omega}_{>}}\right)^{3}\mathcal{F}_{3}(z_{>}),&
N_{<}=
N_{\rm c}^{<}+
\displaystyle{
\left(\frac{k_{\rm B}T}{\hbar\bar{\omega}_{<}}\right)^{3}
}
\mathcal{G}_{3}(z_{<}),&
N_{\rm m}=
\left(\frac{k_{\rm B}T}{\hbar\bar{\omega}_{\rm m}}\right)^{3}\mathcal{F}_{3}(z_{\rm m}),  
\end{eqnarray}
where
$N_{\rm c}^{<}$ is the condensate population of minority atoms. 
The condition for a minority component BEC is given by $\mu_{1}=-\mu_{2}$. 
Fig.~\ref{FBFTc.fig} shows 
the phase diagram 
showing the phase boundary for BEC 
in a bosonic minority component, 
where we assume $\alpha = 3/4$ and 
equal trap frequencies 
$\bar{\omega}_{>} = \bar{\omega}_{<} = \bar{\omega}_{\rm m}$ for simplicity. 
Fig.~\ref{FBFN.fig} shows 
fraction of each component. 
Figs.~\ref{FBFN.fig} (A),~\ref{FBFN.fig} (B) and~\ref{FBFN.fig} (C) show 
fractions of the majority atomic, the minority atomic and the heteronuclear molecular 
components defined by $N_{>}/N_{\rm tot}$, $N_{<}/N_{\rm tot}$ and 
$N_{\rm m}/N_{\rm tot}$. 
Fig.~\ref{FBFN.fig} (D) shows 
the condensed fraction of the minority atomic component defined by 
$N_{\rm c}^{<}/N_{\rm tot}$. 
Fig.~\ref{FBFS.fig} shows the contours of constant entropy 
assuming the system traverses through the adiabatic ramp, 
where the total entropy is given by 
$S = S_{\rm F}(z_{>})+\tilde{S}_{\rm B}(z_{<})+S_{\rm F}(z_{\rm m})$. 

As discussed in the previous section, 
populations and adiabatic trajectories in $\delta$-$T$ plane 
assuming equal populations and different trap frequencies 
are plotted in Ref.~\cite{Morales}. 
Although we use different parameters from ones in Ref.~\cite{Morales}, 
characteristic behaviors do not change. 
Using this system, Ref.~\cite{Morales} proposed cooling cycle with adiabatic sweep.

The critical detuning $\delta_{\rm c}$ 
above which Bose condensation occurs at zero temperature is given by 
\begin{eqnarray}
\delta_{\rm c} = 
\hbar\bar{\omega}_{>}
\left(6\frac{1-\alpha}{1+\alpha}N_{\rm tot}\right)^{1/3}
-
\hbar\bar{\omega}_{\rm m}
\left(6\frac{\alpha}{1+\alpha}N_{\rm tot}\right)^{1/3}. 
\end{eqnarray}
At $\delta = \delta_{\rm c}$ and $T = 0$, 
all atoms in a minority component are converted into heteronuclear molecules 
since there is no bosonic thermal component $\tilde{N}_{<}=0$. 
At this point, 
populations of the majority component and the heteronuclear molecule 
are given by $N_{>}=N_{\rm tot}(1-\alpha)/(1+\alpha)$ and 
$N_{\rm m}=N_{\rm tot}\alpha/(1+\alpha)$. 
The critical value $\delta_{\rm c}$ plays a role of matching 
the Fermi surfaces of a majority component and the heteronuclear molecule, 
{\it i.e.}, $E_{\rm F}^{>} = E_{\rm F}^{\rm m} + \delta_{\rm c}$, 
where $E_{\rm F}^{>}$ and $E_{\rm F}^{\rm m}$ are Fermi energies 
of the fermionic majority component and the fermionic heteronuclear molecule. 
With decreasing 
population of the minority bosonic component, 
the transition temperature of Bose-Einstein condensation $T_{\rm c}$ 
decreases. 
In the limit $\alpha \rightarrow 0$, 
$T_{\rm c}$ goes to zero, 
and the critical detuning $\delta_{\rm c}$ approaches the Fermi energy of the majority component 
$\delta_{\rm c} \rightarrow E_{\rm F}^{>}$. 
Of course, 
the critical detuning $\delta_{\rm c}$ in the case $\alpha = 0$ has no physical meaning, 
because one has no minority bosonic atoms and no heteronuclear molecules. 

We note that the condition for a minority component BEC is given by $\mu_{1}=-\mu_{2}$. 
Under this condition, 
the transition temperature $T_{\rm c}$ 
at zero detuning $\delta = 0$ is given by 
\begin{eqnarray}
k_{\rm B}T_{\rm c}&=&\hbar\bar{\omega}_{>}
\left[\frac{\alpha-(1-\alpha)\gamma_{\rm m}^{3}}
{\gamma_{<}^{3}(1+\gamma_{\rm m}^{3})
\zeta(3)(1+\alpha)}N_{\rm tot}\right]^{1/3}.
\label{48eq}
\end{eqnarray}
From Eq. (\ref{48eq}), 
we find that the minority component 
does not undergo Bose condensation 
at zero detuning 
if the initial number ratio $\alpha$ 
is less than a critical value 
$\alpha_{\rm c} \equiv 
\bar{\omega}_{>}^{3}/(\bar{\omega}_{>}^{3}+\bar{\omega}_{\rm m}^{3})$. 
We note that $\delta_{\rm c}$ is positive for 
$\alpha < \alpha_{\rm c}$.

We first consider the case, 
$\alpha < \alpha_{\rm c}$, 
where no condensation exists at $\delta = 0$. 
This means that all of the atomic minority component are converted into molecules at $T = 0$. 
In such a case, the populations are given by 
\begin{eqnarray}
\left \{
\begin{array}{lll}
N_{>}(\delta = 0)&=&N_{>,{\rm ini}} - N_{<,{\rm ini}} 
= 
\displaystyle {
\frac{1-\alpha}{1+\alpha}
}
N_{\rm tot}, 
\\
N_{<}(\delta = 0)&=&0,
\\
N_{\rm m}(\delta = 0)&=&N_{<,{\rm ini}} = 
\displaystyle { 
\frac{\alpha}{1+\alpha}
}
N_{\rm tot}. 
\end{array}
\right.
\end{eqnarray} 
In this case, the maximum conversion efficiency $\chi_{0}$ 
reaches $100\%$ at $T=0$. 

In the low temperature limit, 
the final entropy at zero detuning $\delta = 0$ is given by 
\begin{eqnarray}
S_{\rm f} &= &S_{\rm F}(z_{>}) + \tilde{S}_{\rm B}(z_{<}) + S_{\rm F}(z_{\rm m}) 
\nonumber
\\
&\approx& 
k_{\rm B}\pi^{2}\left[\frac{(N_{>})^{2}}{6}\right]^{1/3}
\frac{k_{\rm B}T_{\rm f}}{\hbar\bar{\omega}_{>}}
+ 4k_{\rm B}\left(\frac{k_{\rm B}T_{\rm f}}{\hbar\bar{\omega}_{<}}\right)^{3}\zeta(4)
+ k_{\rm B}\pi^{2}\left[\frac{(N_{\rm m})^{2}}{6}\right]^{1/3}
\frac{k_{\rm B}T_{\rm f}}{\hbar\bar{\omega}_{\rm m}}
\nonumber
\\
&\approx&
k_{\rm B}\pi^{2}
\left ( \frac{1}{6} \right )^{1/3}
\left [ 
\left ( 
\frac{1-\alpha}{1+ \alpha}N_{\rm tot}
\right )^{2/3}
+ 
\gamma_{\rm m} 
\left ( 
\frac{\alpha}{1+ \alpha}N_{\rm tot}
\right )^{2/3}
\right ]
\frac{k_{\rm B}T_{\rm f}}{\hbar\bar{\omega}_{>}}, 
\label{47eq}
\end{eqnarray}
where the approximation leading to the last line is valid when 
the final temperature satisfies the condition 
\begin{eqnarray}
k_{\rm B}T_{\rm f} \ll 
\sqrt{\frac{1}{4\zeta (4)}}\pi
\left ( \frac{1}{6} \right )^{1/6}
\left ( \frac{1}{1+\alpha} N_{\rm tot}\right)^{1/3}
\sqrt{(1-\alpha)^{2/3}+\alpha^{2/3}}
\frac{1}{\gamma_{<}^{3/2}}
\hbar\bar{\omega}_{>} \,\,. 
\end{eqnarray}

In a pure atomic gas before a sweep of the magnetic field, 
the initial populations are given by 
\begin{eqnarray}
\left \{
\begin{array}{lllll}
N_{>,{\rm ini}}&=&
\displaystyle{
\frac{1}{1+\alpha}
}
N_{\rm tot}&=&
\displaystyle{
\left(\frac{k_{\rm B}T_{\rm ini}}{\hbar\bar{\omega}_{>}}\right)^{3}
}
\mathcal{F}_{3}(z_{>,{\rm ini}}), 
\\
N_{<,{\rm ini}}&=&
\displaystyle{
\frac{\alpha}{1+\alpha}N_{\rm tot}
}
&=&
N_{\rm c, ini}^{<} + 
\displaystyle{
\left(\frac{k_{\rm B}T_{\rm ini}}{\hbar\bar{\omega}_{<}}\right)^{3}
}
\mathcal{G}_{3}(z_{<,{\rm ini}}). 
\end{array}
\right.
\end{eqnarray} 
The initial transition temperature $T_{\rm c, ini}$ and 
a condensate population $N_{\rm c, ini}^{<}$ below $T_{\rm c, ini}$ are given by 
\begin{eqnarray}
\left \{
\begin{array}{lll}
k_{\rm B}T_{\rm c, ini}&=&\hbar\bar{\omega}_{<}
\displaystyle {
\left(\frac{N_{<,{\rm ini}}}{\zeta(3)}\right)^{1/3}
}, 
\\
N_{\rm c, ini}^{<}&=&
\displaystyle { 
\frac{\alpha}{1+\alpha}N_{\rm tot}
}
-
\displaystyle { 
\left(\frac{k_{\rm B}T_{\rm ini}}{\hbar\bar{\omega}_{<}}\right)^{3}\zeta(3). 
}
\end{array}
\right.
\end{eqnarray}
In the low temperature limit, 
the initial entropy is given by 
\begin{eqnarray}
S_{\rm ini} &= &S_{\rm F}(z_{>,{\rm ini}}) + S_{\rm B}(z_{<,{\rm ini}})
\nonumber
\\
&\approx& 
k_{\rm B}\pi^{2}\left[\frac{(N_{>,{\rm ini}})^{2}}{6}\right]^{1/3}
\frac{k_{\rm B}T_{\rm ini}}{\hbar\bar{\omega}_{>}}
+4k_{\rm B}\left(\frac{k_{\rm B}T_{\rm ini}}{\hbar\bar{\omega}_{<}}\right)^{3}\zeta(4)
\nonumber
\\
&\approx&
k_{\rm B}\pi^{2}\left[\frac{1}{6}
\left(\frac{1}{1+\alpha}N_{\rm tot}\right)^{2}\right]^{1/3}
\frac{k_{\rm B}T_{\rm ini}}{\hbar\bar{\omega}_{>}}, 
\label{51eq}
\end{eqnarray}
where the approximation leading to the last line is valid when 
the initial temperature satisfies the condition 
\begin{eqnarray}
k_{\rm B}T_{\rm ini} \ll 
\sqrt{\frac{1}{4\zeta (4)}}\pi
\left ( \frac{1}{6} \right )^{1/6}
\left ( \frac{1}{1+\alpha} N_{\rm tot} \right)^{1/3}
\frac{1}{\gamma_{<}^{3/2}}
\hbar\bar{\omega}_{>}. 
\end{eqnarray}

\begin{figure}
\begin{center}
\includegraphics[width=16cm,height=6cm,keepaspectratio,clip]{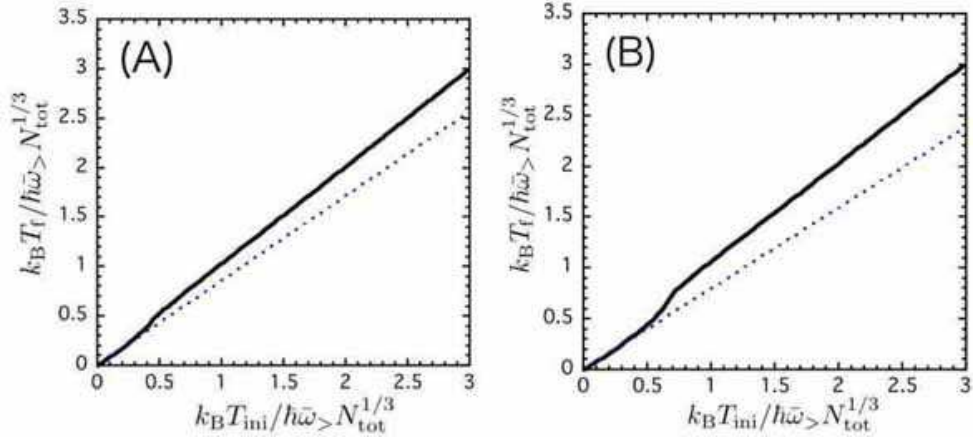}
\end{center}
\caption{
(Color online) 
The relation between the initial temperature and the final temperature at zero detuning $\delta = 0$. 
We assume equal trap frequencies 
$\bar{\omega}_{>} = \bar{\omega}_{<} = \bar{\omega}_{\rm m}$. 
In this case, a critical value $\alpha_{\rm c} =1/2$ becomes a boundary, 
above which we have the condensation of the atomic minority component 
at zero detuning $\delta = 0$. 
Fig. (A) shows the relation with $\alpha=2/15 (< \alpha_{\rm c} =1/2)$ 
and Fig. (B) shows the relation with $\alpha = 3/4 (> \alpha_{\rm c} =1/2)$. 
Solid lines are numerical results and 
dotted lines are results of low temperature limit. 
}
\label{FBFTfTi.fig}
\end{figure}

Connecting the final entropy at zero detuning $\delta = 0$ with the initial entropy, 
we obtain the relation between the initial temperature $T_{\rm ini}$ and 
the final temperature $T_{\rm f}$. In the low temperature limit, 
using Eqs. (\ref{47eq}) and (\ref{51eq}), 
we obtain the explicit formula 
\begin{eqnarray}
T_{\rm f}=\frac{1}{(1-\alpha)^{2/3}+\alpha^{2/3}}T_{\rm ini}. 
\label{TfTiLowFBFnoBEC}
\end{eqnarray}
In the high temperature limit, 
the heteronuclear molecular population is so small 
that the contribution of the heteronuclear molecule entropy to the total entropy is small. 
As a result, one has $T_{\rm f} \approx T_{\rm ini}$ in the high temperature region. 
Fig.~\ref{FBFTfTi.fig} (A) shows the numerical result for 
the final temperature as a function of the initial temperature, assuming 
$\alpha = 2/15$ 
(which is less than 
$\alpha_{\rm c} = \bar{\omega}_{>}^{3}/(\bar{\omega}_{>}^{3}+\bar{\omega}_{\rm m}^{3})=1/2$
for 
$\bar{\omega}_{>} = \bar{\omega}_{<} = \bar{\omega}_{\rm m}$). 
We also plot the result of 
the low temperature limit given in Eq. (\ref{TfTiLowFBFnoBEC}).

We next consider the case 
$\alpha\geq \alpha_{\rm c}=\bar{\omega}_{>}^{3}/(\bar{\omega}_{>}^{3}+\bar{\omega}_{\rm m}^{3})$, 
where a condensate exists 
at zero detuning $\delta = 0$. 
The condensate population of the atomic minority component 
at zero detuning $\delta = 0$ is given by 
\begin{eqnarray}
N_{\rm c}^{<} = 
\displaystyle{
\frac{(1+\gamma_{\rm m}^{3})\alpha-\gamma_{\rm m}^{3}}
{(1+\alpha)(1+\gamma_{\rm m}^{3})}N_{\rm tot}
}
-
\displaystyle{
\left ( \frac{k_{\rm B}T}{\hbar\bar{\omega}_{>}} \right)^{3}\zeta(3)
\gamma_{<}^{3}
}. 
\end{eqnarray}
The population of each component below $T_{\rm c}$ 
at zero dutuning $\delta = 0$ is given by 
\begin{eqnarray}
\left \{
\begin{array}{lll}
N_{>}(\delta = 0)&=&N_{>,{\rm ini}}-N_{\rm m}=
\displaystyle{
\frac{1}{1+\alpha}\frac{1}{1+\gamma_{\rm m}^{3}}
}
N_{\rm tot}, 
\\
N_{<}(\delta = 0)&=&
\displaystyle{
\frac{(1+\gamma_{\rm m}^{3})\alpha-\gamma_{\rm m}^{3}}
{(1+\alpha)(1+\gamma_{\rm m}^{3})}
}N_{\rm tot}, 
\\
N_{\rm m}(\delta = 0)&=&N_{<,{\rm ini}}-N_{<}=
\displaystyle{
\frac{1}{1+\alpha}\frac{\gamma_{\rm m}^{3}}{1+\gamma_{\rm m}^{3}}
}
N_{\rm tot}. 
\end{array}
\right .
\end{eqnarray}
The population ratio of a fermionic majority component and a fermionic hetorunuclear molecule are constant 
with the bosonic minority component being Bose condensed, 
consistent with the fugacity relation $z_{>}=z_{\rm m}$ at zero detuning $\delta = 0$ 
below $T_{\rm c}$ where $z_{<}=1$. 

The final entropy at zero detuning $\delta=0$ in the low temperature limit is given by 
\begin{eqnarray}
S_{\rm f} &= &S_{\rm F}(z_{>,{\rm f}}) + \tilde{S}_{\rm B}(z_{<,{\rm f}}) + S_{\rm F}(z_{\rm m, f})
\nonumber
\\
&\approx& 
k_{\rm B}\pi^{2}\left[\frac{(N_{>})^{2}}{6}\right]^{1/3}
\frac{k_{\rm B}T_{\rm f}}{\hbar\bar{\omega}_{>}}
+ 4k_{\rm B}\left(\frac{k_{\rm B}T_{\rm f}}{\hbar\bar{\omega}_{<}}\right)^{3}\zeta(4)
+ k_{\rm B}\pi^{2}\left[\frac{(N_{\rm m})^{2}}{6}\right]^{1/3}
\frac{k_{\rm B}T_{\rm f}}{\hbar\bar{\omega}_{\rm m}}
\nonumber
\\
&\approx&
2k_{\rm B}\pi^{2}\left ( \frac{1}{6} \right )^{3}
\left ( \frac{1}{1+\alpha}N_{\rm tot} \right )^{2/3}
(1+\gamma_{\rm m}^{3})^{1/3}
\frac{k_{\rm B}T_{\rm f}}{\hbar\bar{\omega}_{>}}, 
\end{eqnarray}
where the approximation leading to the last line is valid for 
\begin{eqnarray}
k_{\rm B}T_{\rm f} \ll 
\sqrt{\frac{1}{4\zeta (4)}}\pi
\left ( \frac{1}{6} \right )^{1/6}
\left ( \frac{1}{1+\alpha} N_{\rm tot} \right)^{1/3}
(1+\gamma_{\rm m}^{3})^{1/6}
\frac{1}{\gamma_{<}^{3/2}}
\hbar\bar{\omega}_{>}. 
\end{eqnarray}

By connecting the final entropy with the initial entropy, 
we obtain the relation between the initial temperature 
and the final temperature. 
In Fig.~\ref{FBFTfTi.fig} (B), 
we plot the numerical result for the final temperature 
assuming $\alpha = 3/4$ 
(which is greater than $\alpha_{\rm c} = \bar{\omega}_{>}^{3}/(\bar{\omega}_{>}^{3}+\bar{\omega}_{\rm m}^{3})=1/2$ 
for 
$\bar{\omega}_{>} = \bar{\omega}_{<}=\bar{\omega}_{\rm m}$). 
We also plot the analytical result in the low temperature limit, 
given by 
\begin{eqnarray}
T_{\rm f}=\left(\frac{1}{1+\gamma_{\rm m}^{3}}\right)^{1/3}T_{\rm ini}. 
\end{eqnarray}

Below $T_{\rm c}$ and at zero detuning $\delta = 0$, 
the population ratio of the majority component and the heteronuclear molecule 
becomes constant, 
which follows from the fugacity relation $z_{>} = z_{\rm m}$. 
As a result, the molecular conversion efficiency below $T_{\rm c}$ is given by 
\begin{eqnarray}
\chi_{0} = 
\frac{1}{\alpha} 
\frac{\bar{\omega}_{>}^{3}}{\bar{\omega}_{>}^{3}+\bar{\omega}_{\rm m}^{3}}.
\label{59eq}
\end{eqnarray}

\begin{figure}
\begin{center}
\includegraphics[width=8cm,height=8cm,keepaspectratio,clip]{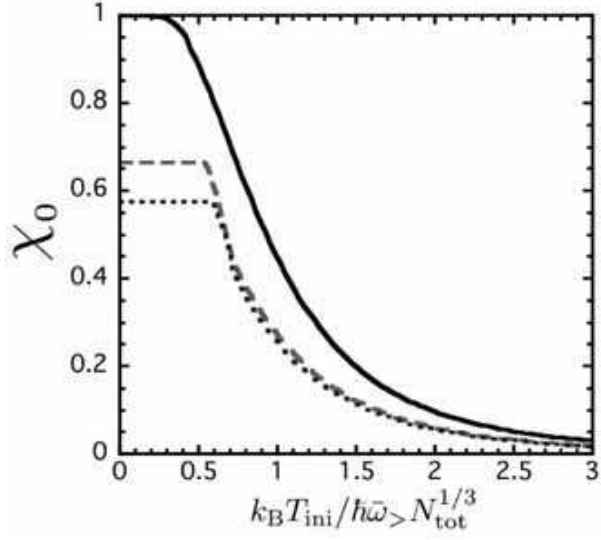}
\end{center}
\caption{
(Color online) 
The molecular conversion efficiency 
as a function of the initial temperature. 
We assume equal trap frequencies 
$\bar{\omega}_{>} = \bar{\omega}_{<} = \bar{\omega}_{\rm m}$. 
In this case, a critical value $\alpha_{\rm c}=1/2$ becomes a boundary, 
above which we have the condensation of the atomic minority component 
at zero detuning $\delta = 0$. 
We assume the ratio $\alpha = 2/15$(solid line), 
$\alpha = 3/4$(dashed line) 
and $\alpha = 13/15$(dotted line). 
}
\label{FBFchi.fig}
\end{figure}

Fig.~\ref{FBFchi.fig} shows the conversion efficiency 
as a function of the initial temperature. 
The solid line shows the conversion efficiency 
for $\alpha = 2/15$, 
where Bose condensation does not occur. 
The dashed and dotted lines show the conversion efficiencies 
for $\alpha = 3/4$ and $13/15$, respectively, 
where a minority atomic gas becomes Bose-condensed at low temperature. 
A critical value $\alpha_{\rm c} =  \bar{\omega}_{>}^{3}/(\bar{\omega}_{>}^{3}+\bar{\omega}_{\rm m}^{3})$ 
($= 1/2$ 
for 
$\bar{\omega}_{>} =\bar{\omega}_{<} =\bar{\omega}_{\rm m}$)
is a boundary 
that makes the trend of the conversion efficiency change, 
as shown in Fig.~\ref{FBFchi.fig}. 

\begin{figure}
\begin{center}
\includegraphics[width=8cm,height=8cm,keepaspectratio,clip]{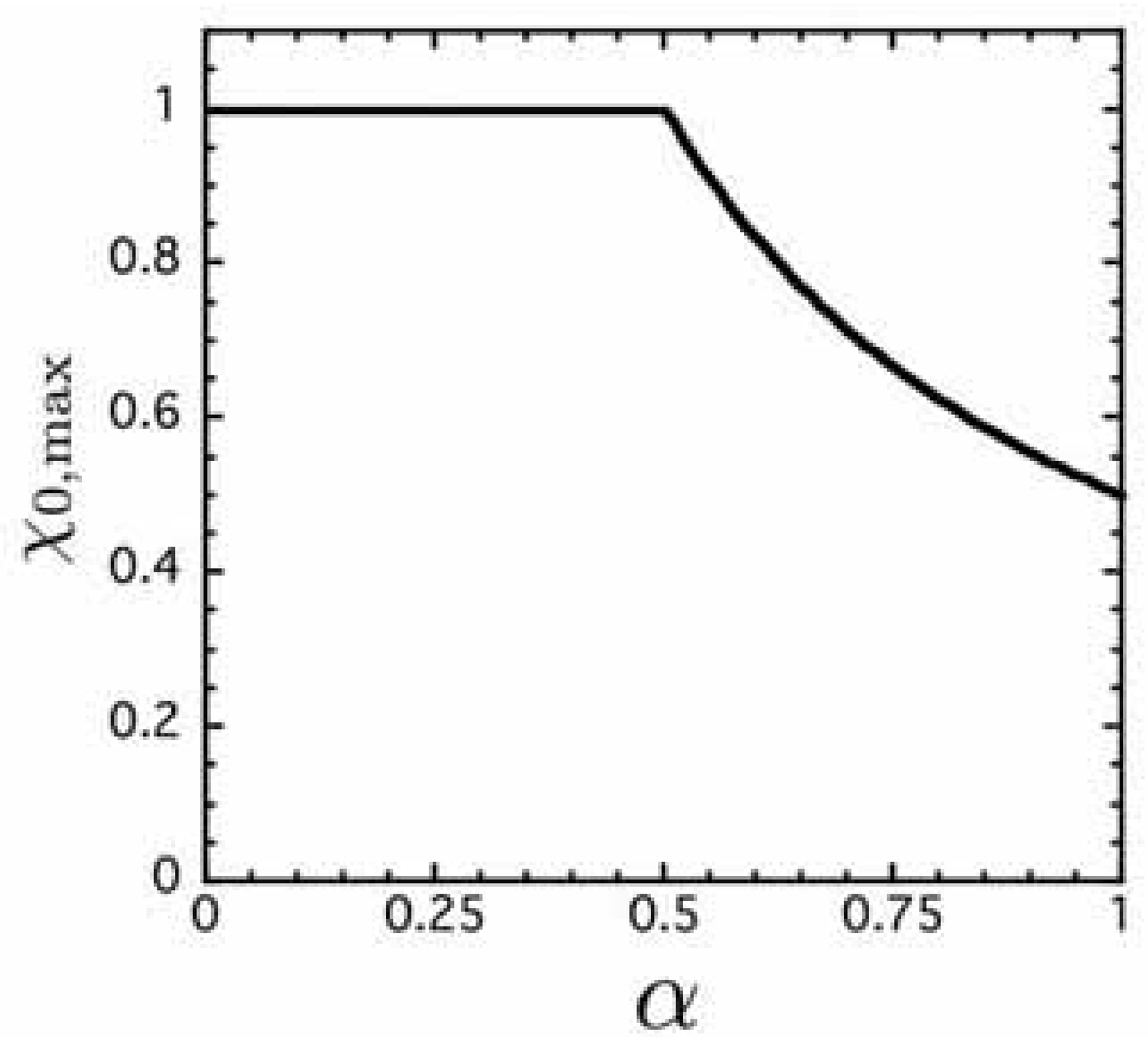}
\end{center}
\caption{
The maximum molecular conversion efficiency $\chi_{0, {\rm max}}$
as a function of the initial population ratio of atoms $\alpha$. 
We assume equal trap frequencies 
$\bar{\omega}_{>} = \bar{\omega}_{<} = \bar{\omega}_{\rm m}$. 
}
\label{FBFchiMAX.fig}
\end{figure}

In Fig.~\ref{FBFchiMAX.fig}, 
we plot the $\alpha$-dependence of the maximum conversion efficiency 
for equal trap frequencies 
$\bar{\omega}_{>} = \bar{\omega}_{<} = \bar{\omega}_{\rm m}$. 
The maximum conversion efficiency $\chi_{0, {\rm max}}$ 
dramatically changes at $\alpha = 1/2$. 
The maximum conversion efficiency $\chi_{0, {\rm max}}$ ranges 
from $50\%$ to $100\%$, 
depending on the initial number ratio $\alpha$. 
In general, for $\alpha \geq \alpha_{\rm c} = \bar{\omega}_{>}^{3}/(\bar{\omega}_{>}^{3}+\bar{\omega}_{\rm m}^{3})$, 
the maximum conversion is given by 
$\bar{\omega}_{>}^{2}/[\alpha(\bar{\omega}_{>}^{3}+\bar{\omega}_{\rm m}^{3})]$ 
with a plateau. 
The maximum conversion efficiency ranges 
from $\bar{\omega}_{>}^{3}/(\bar{\omega}_{>}^{3}+\bar{\omega}_{\rm m}^{3})$ 
to unity, as a function of the initial atomic fraction $\alpha$. 
This result would not be expected 
if one applied the SPSS model as discussed in Sec.~\ref{SecBFF}.

Two cases $\{{\rm B}_{>}+{\rm F}_{<}\leftrightarrow {\rm F}_{\rm m}\}$ and 
$\{{\rm F}_{>}+{\rm B}_{<}\leftrightarrow {\rm F}_{\rm m}\}$ are essentially the same. 
In the case $\{{\rm B}_{>}+{\rm F}_{<}\leftrightarrow {\rm F}_{\rm m}\}$, 
maximum conversion efficiency with a plateau is given by 
$\chi_{0, {\rm max}}=\bar{\omega}_{\rm F}^{3}
/(\bar{\omega}_{\rm F}^{3}+\bar{\omega}_{\rm m}^{3})$, 
where a trap frequency of the fermionic minority component 
$\bar{\omega}_{<}$ is denoted as $\bar{\omega}_{\rm F}$. 
On the other hand, in the case $\{{\rm F}_{>}+{\rm B}_{<}\leftrightarrow {\rm F}_{\rm m}\}$, 
maximum conversion efficiency for 
$\alpha > \alpha_{\rm c} =\bar{\omega}_{\rm F}^{3}/
(\bar{\omega}_{\rm F}^{3}+\bar{\omega}_{\rm m}^{3})$ 
is given by 
$\chi_{0, {\rm max}}=\bar{\omega}_{\rm F}^{3}
/[\alpha(\bar{\omega}_{\rm F}^{3}+\bar{\omega}_{\rm m}^{3})]$, 
where a trap frequency of the fermionic majority component 
$\bar{\omega}_{>}$ is denoted as $\bar{\omega}_{\rm F}$. 
We find that the maximum conversion efficiency is determined 
by initial number ratio and trap frequencies of fermionic components, 
which are both atoms and heteronuclear molecules. 
This result shows that fermionic components play a crucial role in determining 
each population, 
because energies of fermionic components are dominant 
for the total energy rather than bosonic one 
in the low temperature region. 
We note that 
one can achieve large conversion efficiency 
by making the trap frequencies of atoms larger than that of heteronuclear molecules, 
because the density of states of the heteronuclear molecule is higher than that of atoms in this case. 
This becomes more apparent when one writes 
the maximum conversion efficiency 
in Eq. (\ref{59eq}) as 
\begin{eqnarray}
\displaystyle{
\frac{1}{\alpha}
\frac{\bar{\omega}_{\rm a}^{3}}
{(\bar{\omega}_{\rm a}^{3}+\bar{\omega}_{\rm m}^{3})}
}
= 
\displaystyle{
\frac{1}{2\alpha}
}
+
\displaystyle{
\frac{\bar{\omega}_{\rm a}^{3}-\bar{\omega}_{\rm m}^{3}}
{2\alpha(\bar{\omega}_{\rm a}^{3}+\bar{\omega}_{\rm m}^{3})}
},
\end{eqnarray}
where we define an atomic trap frequencies as $\bar{\omega}_{\rm a}$. 
This effect can be also seen in the case 
$\{{\rm B}_{>}+{\rm B}_{<}\leftrightarrow {\rm B}_{\rm m}\}$, 
as well as cases 
$\{{\rm B}_{>}+{\rm F}_{<}\leftrightarrow {\rm F}_{\rm m}\}$ and 
$\{{\rm F}_{>}+{\rm B}_{<}\leftrightarrow {\rm F}_{\rm m}\}$.

\section{Gas of Majority Fermionic Atom, Minority Fermionic Atom 
and its Heteronuclear Molecule  \label{SecFFB}}
\begin{figure}
\begin{center}
\includegraphics[width=8cm,height=8cm,keepaspectratio,clip]{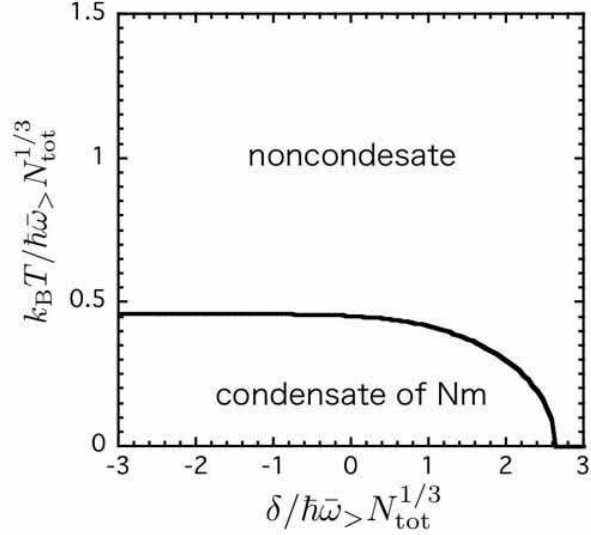}
\end{center}
\caption{
The phase diagram with the transition temperature 
of a bosonic minority component. 
We assume the ratio $\alpha = 2/15$, 
and equal trap frequencies 
$\bar{\omega}_{>} = \bar{\omega}_{<} = \bar{\omega}_{\rm m}$. 
}
\label{FFBTc.fig}
\end{figure}

\begin{figure}
\begin{center}
\includegraphics[width=15cm,height=15cm,keepaspectratio,clip]{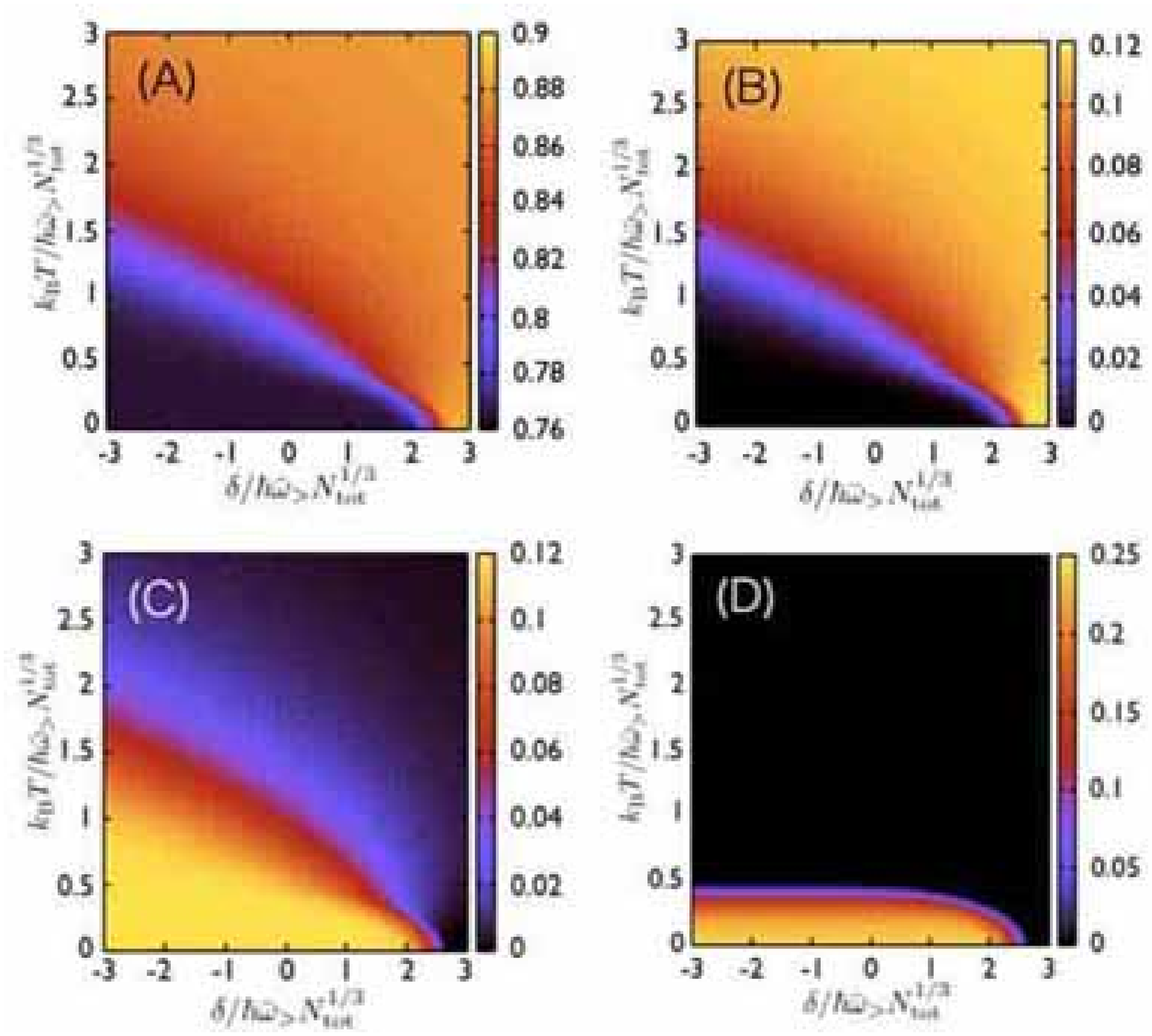}
\end{center}
\caption{
(Color online) 
(A) The fraction of a fermionic majority component: $N_{>}/N_{\rm tot}$. 
(B) The fraction of a fermionic minority component: $N_{<}/N_{\rm tot}$. 
(C) The fraction of a bosonic heteronuclear molecular component: $N_{\rm m}/N_{\rm tot}$. 
(D) The condensate fraction of a bosonic heteronuclear molecular component: $N_{\rm c}^{\rm m}/N_{\rm tot}$. 
We assume the ratio $\alpha = 2/15$, 
and equal trap frequencies 
$\bar{\omega}_{>} = \bar{\omega}_{<} = \bar{\omega}_{\rm m}$. 
}
\label{FFBN.fig}
\end{figure}

\begin{figure}
\begin{center}
\includegraphics[width=8cm,height=8cm,keepaspectratio,clip]{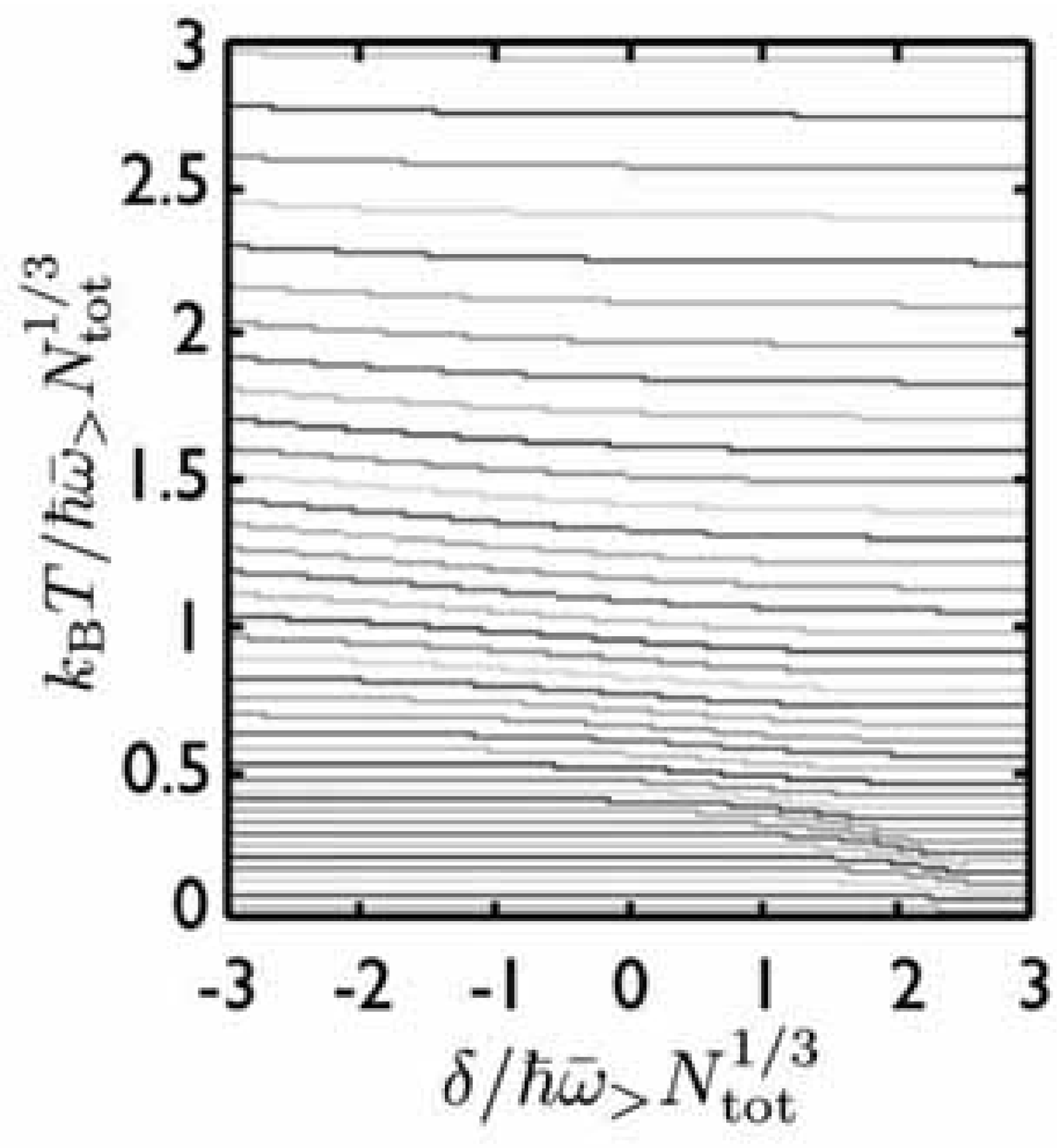}
\end{center}
\caption{
(Color online) 
The contours of constant entropy. 
We assume the ratio $\alpha = 2/15$, 
and equal trap frequencies 
$\bar{\omega}_{>} = \bar{\omega}_{<} = \bar{\omega}_{\rm m}$. 
}
\label{FFBS.fig}
\end{figure}

Finally, we consider the case with a majority fermionic atom: $\{{\rm F}_{>}+{\rm F}_{<}\leftrightarrow {\rm B}_{\rm m}\}$. 
Under the two constraints in Eqs. (\ref{eq2}) and (\ref{eq4}), 
the populations are given by 
\begin{eqnarray}
N_{>}=
\displaystyle{
\left(\frac{k_{\rm B}T}{\hbar\bar{\omega}_{>}}\right)^{3}
}
\mathcal{F}_{3}(z_{>}),&
N_{<}=
\displaystyle{
\left(\frac{k_{\rm B}T}{\hbar\bar{\omega}_{<}}\right)^{3}
}
\mathcal{F}_{3}(z_{<}),&
N_{\rm m}=
N_{\rm c}^{\rm m} + 
\displaystyle{
\left(\frac{k_{\rm B}T}{\hbar\bar{\omega}_{\rm m}}\right)^{3}
}
\mathcal{G}_{3}(z_{\rm m}),  
\end{eqnarray}
where
$N_{\rm c}^{\rm m}$ is the population of the condensed heteronuclear molecule. 
The condition for heteronuclear molecular condensation is 
$2\mu_{1}+(1-\alpha)\mu_{2}-\delta=0$. 
Fig.~\ref{FFBTc.fig} shows 
the phase diagram with the transition temperature where 
the heteronuclear molecules become Bose condensed. 
We assume $\alpha = 2/15$, 
and equal trap frequencies 
$\bar{\omega}_{>} = \bar{\omega}_{<} = \bar{\omega}_{\rm m}$. 
Fig.~\ref{FFBN.fig} shows fraction of each component. 
Figs.~\ref{FFBN.fig} (A),~\ref{FFBN.fig} (B) and~\ref{FFBN.fig} (C) show fractions of the majority atomic, 
the minority atomic and the molecular components 
defined by $N_{>}/N_{\rm tot}$, $N_{<}/N_{\rm tot}$ and $N_{\rm m}/N_{\rm tot}$. 
Fig.~\ref{FFBN.fig} (D) shows 
the heteronuclear molecule condensate fraction defined by $N_{\rm c}^{\rm m}/N_{\rm tot}$. 
Fig.~\ref{FFBS.fig} shows the contours of constant entropy 
assuming the system traverses through the adiabatic ramp, 
where the total entropy is given by 
$S = S_{\rm F}(z_{>})+S_{\rm F}(z_{<})+\tilde{S}_{\rm B}(z_{\rm m})$.

In Refs.~\cite{watabe:2006} and~\cite{williams2004adiabatic}, 
populations and adiabatic trajectories assuming population balanced gas 
and equal trap frequencies are plotted in $\delta$-$T$ plane. 
The characteristic behaviors plotted in this section 
do not change compared with results 
in Refs.~\cite{watabe:2006} and~\cite{williams2004adiabatic}.

The critical detuning $\delta_{\rm c}$ 
below which Bose condensation of molecules 
occurs at zero temperature $T=0$ is given by 
\begin{eqnarray}
\delta_{\rm c} = 
\hbar\bar{\omega}_{>}
\left(6\frac{1}{1+\alpha}N_{\rm tot}\right)^{1/3}
+
\hbar\bar{\omega}_{<}
\left(6\frac{\alpha}{1+\alpha}N_{\rm tot}\right)^{1/3}. 
\end{eqnarray}
At $\delta = \delta_{\rm c}$,  
heteronuclear molecules do not exist since there is no thermal component at $T=0$. 
At this point, 
populations of the majority and the minority components 
are given by $N_{>}=N_{\rm tot}/(1+\alpha)$ and 
$N_{<}=N_{\rm tot}\alpha/(1+\alpha)$. 
The critical value $\delta_{\rm c}$ is also represented as 
$\delta_{\rm c} = E_{\rm F}^{>} + E_{\rm F}^{<}$, 
where $E_{\rm F}^{>}$ and $E_{\rm F}^{<}$ are Fermi energies 
of a majority and minority components. 
Just below $\delta_{\rm c}$ at $T=0$, 
although the single-particle lowest energy of heteronuclear molecules is higher than that of atoms, 
condensed heteronuclear molecules exist owing to Pauli's exclusion principle.

At zero detuning $\delta = 0$, 
the condition for heteronuclear molecular condensation is given by 
$\mu_{1}=-(1-\alpha)\mu_{2}/2$. 
Chemical potentials of majority and minority components 
$\mu_{>}$ and $\mu_{<}$ are given by 
$\mu_{>} = -(1+\alpha)\mu_{2}$ and $\mu_{<} = (1+\alpha)\mu_{2}$. 
At $T=0$, 
a part of atoms in a majority component is not converted into heteronuclear molecules, 
which forms the Fermi surface. 
From this condition, the chemical potential of a majority component 
should be positive, i.e. $\mu_{2}\leq 0$, 
and thus one has no minority component 
due to the negative chemical potential 
at $T=0$ and $\delta = 0$. 
The populations at $T=0$ and zero detuning $\delta =0$ are given by 
\begin{eqnarray}
\left \{
\begin{array}{lll}
N_{>}(\delta = 0)&=&N_{>,{\rm ini}} - N_{<,{\rm ini}} 
= 
\displaystyle{
\frac{1-\alpha}{1+\alpha}
}
N_{\rm tot},
\\
N_{<}(\delta = 0)&=&0,
\\
N_{\rm m}(\delta = 0)&=&N_{<,{\rm ini}} = 
\displaystyle{
\frac{\alpha}{1+\alpha}
}
N_{\rm tot}.
\end{array}
\right.
\end{eqnarray}
At $T=0$, the conversion efficiency $\chi_{0} $ is $100\%$, 
which gives the maximum conversion.

The final entropy at zero detuning $\delta = 0$
in the low temperature limit is given by 
\begin{eqnarray}
S_{\rm f} &= &S_{\rm F}(z_{>}) + S_{\rm F}(z_{<}) + \tilde{S}_{\rm B}(Z_{\rm m})
\nonumber
\\
&\approx& 
k_{\rm B}\pi^{2}\left[\frac{(N_{>})^{2}}{6}\right]^{1/3}
\frac{k_{\rm B}T_{\rm f}}{\hbar\bar{\omega}_{>}}
+
k_{\rm B}\pi^{2}\left[\frac{(N_{<})^{2}}{6}\right]^{1/3}
\frac{k_{\rm B}T_{\rm f}}{\hbar\bar{\omega}_{<}}
+ 4k_{\rm B}\left(\frac{k_{\rm B}T_{\rm f}}
{\hbar\bar{\omega}_{\rm m}}\right)^{3}\zeta(4)
\nonumber
\\
&\approx&
k_{\rm B}\pi^{2}\left[\frac{1}{6}
\left(\frac{1-\alpha}{1+\alpha}N_{\rm tot}\right)^{2}\right]^{1/3}
\frac{k_{\rm B}T_{\rm f}}{\hbar\bar{\omega}_{>}}
+ 4k_{\rm B}\left(\frac{k_{\rm B}T_{\rm f}}
{\hbar\bar{\omega}_{\rm m}}\right)^{3}\zeta(4) 
\nonumber
\\
&\approx&
k_{\rm B}\pi^{2}\left[\frac{1}{6}
\left(\frac{1-\alpha}{1+\alpha}N_{\rm tot}\right)^{2}\right]^{1/3}
\frac{k_{\rm B}T_{\rm f}}{\hbar\bar{\omega}_{\rm m}}, 
\label{64eq}
\end{eqnarray}
where the approximation leading to the last line is valid for 
\begin{eqnarray}
k_{\rm B}T_{\rm f} \ll
\sqrt{\frac{1}{4\zeta (4)}}\pi 
\left ( \frac{1}{6} \right )^{1/6}
\left ( \frac{1-\alpha}{1+\alpha} N_{\rm tot}\right )^{1/3} 
\frac{1}{\gamma_{\rm m}^{3/2}}
\hbar\bar{\omega}_{>}. 
\end{eqnarray}
On the other hand, 
the initial entropy in the low temperature limit is given by 
\begin{eqnarray}
S_{\rm ini} &= &S_{>,{\rm ini}} + S_{<,{\rm ini}}
\nonumber
\\
&\approx& 
k_{\rm B}\pi^{2}\left[\frac{(N_{>,{\rm ini}})^{2}}{6}\right]^{1/3}
\frac{k_{\rm B}T_{\rm ini}}{\hbar\bar{\omega}_{>}}
+
k_{\rm B}\pi^{2}\left[\frac{(N_{<,{\rm ini}})^{2}}{6}\right]^{1/3}
\frac{k_{\rm B}T_{\rm ini}}{\hbar\bar{\omega}_{<}}
\nonumber
\\
&\approx&
k_{\rm B}\pi^{2}\left[\frac{1}{6}
\left(\frac{1}{1+\alpha}N_{\rm tot}\right)^{2}\right]^{1/3}
\frac{k_{\rm B}T_{\rm ini}}{\hbar\bar{\omega}_{>}}
+k_{\rm B}\pi^{2}\left[\frac{1}{6}
\left(\frac{\alpha}{1+\alpha}N_{\rm tot}\right)^{2}\right]^{1/3}
\gamma_{<}
\frac{k_{\rm B}T_{\rm ini}}{\hbar\bar{\omega}_{>}}. 
\label{66eq}
\end{eqnarray}

\begin{figure}
\begin{center}
\includegraphics[width=8cm,height=8cm,keepaspectratio,clip]{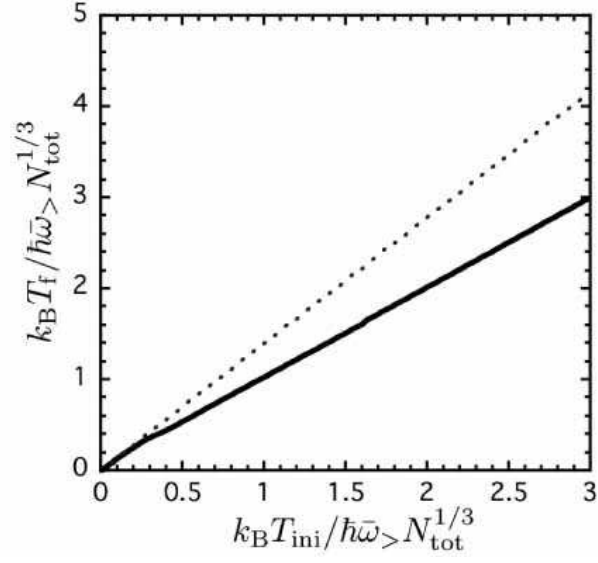}
\end{center}
\caption{
(Color online) 
The relation between the initial temperature and the final temperature 
at zero detuning $\delta = 0$. 
The solid line shows the result of the numerical calculation. 
The dotted line shows the result of the low temperature limit. 
We assume the ratio $\alpha = 2/15$, 
and equal trap frequencies 
$\bar{\omega}_{>} = \bar{\omega}_{<} = \bar{\omega}_{\rm m}$. 
}
\label{FFBTfTi.fig}
\end{figure}

Connecting the final entropy at zero detuning with the initial entropy, 
we obtain the relation between the initial temperature $T_{\rm ini}$ and 
the final temperature $T_{\rm f}$. 
In the low temperature limit, 
the analytical expression for this relation is given from Eqs. (\ref{64eq}) and (\ref{66eq}) as 
\begin{eqnarray}
T_{\rm f}=
\frac{1+\alpha^{2/3}\gamma_{<}}{(1-\alpha)^{2/3}}T_{\rm ini}. 
\label{TfTiLowFFB}
\end{eqnarray}
In the high temperature limit, 
the heteronuclear molecular population is so small 
that the contribution of the heteronuclear molecule entropy to the total entropy is small. 
As a result, one has $T_{\rm f} \approx T_{\rm ini}$ in the high temperature region. 
Fig.~\ref{FFBTfTi.fig} shows the numerical result for 
the final temperature as a function of the initial temperature for $\alpha = 2/15$. 
We also show the analytical result in the low temperature limit 
given in Eq. (\ref{TfTiLowFFB}). 
In the population balanced case $\alpha = 1$, 
the final entropy in the low temperature limit is only 
composed of the heteronuclear molecule entropy given by 
$S_{\rm f} = 4k_{\rm B}\gamma_{\rm m}^{3}
\left(k_{\rm B}T_{\rm f}/\hbar\bar{\omega}_{>}\right)^{3}\zeta(4)$. 
In this case, the relation between the initial and final temperatures in the low temperature limit is given by 
\begin{eqnarray}
k_{\rm B}T_{\rm f} =
\frac{1}{2}\frac{1}{\gamma_{\rm m}}
\left[\frac{\pi^{2}}{\zeta(4)3^{1/3}} 
(1+\gamma_{<})
k_{\rm B}T_{\rm ini}
\left( \hbar\bar{\omega}_{>} N_{\rm tot}^{1/3} \right)^{2}
\right]^{1/3}. 
\end{eqnarray}

\begin{figure}
\begin{center}
\includegraphics[width=8cm,height=8cm,keepaspectratio,clip]{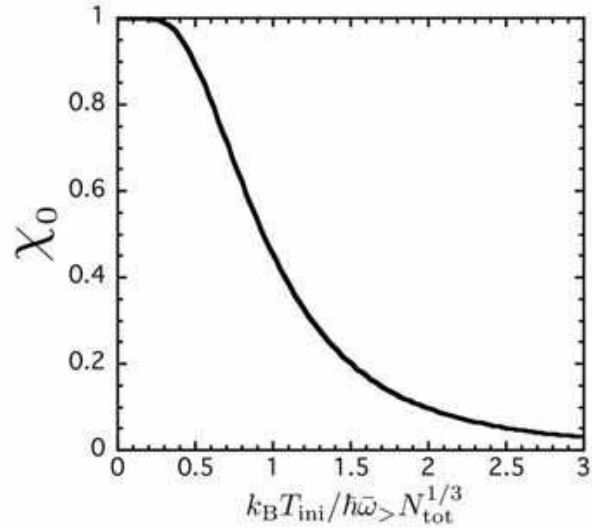}
\end{center}
\caption{
The molecular conversion efficiency 
as a function of the initial temperature. 
We assume the ratio $\alpha = 2/15$, 
and equal trap frequencies 
$\bar{\omega}_{>} = \bar{\omega}_{<} = \bar{\omega}_{\rm m}$. 
}
\label{FFBchi.fig}
\end{figure}

Fig.~\ref{FFBchi.fig} shows the conversion efficiency 
as a function of the initial temperature 
given from the heteronuclear molecular population at zero detuning $\delta = 0$. 
One can see that the maximum conversion reaches $100\%$ at $T = 0$. 
Molecular conversion efficiencies in trapped gases 
composed of two component Fermi atoms 
assuming equal trap frequencies and equal populations 
are discussed in Refs.~\cite{williams:2006} and~\cite{watabe:2006}. 
The behavior shown in Fig.~\ref{FFBchi.fig} is qualitatively consistent 
with the results in Ref.~\cite{watabe:2006}.

\section{Explicit Formula for Conversion Efficiency} 
In Ref.~\cite{hodby:2005}, 
the SPSS model predicted the conversion efficiency 
as a function of the peak phase space density. 
On the other hand, 
Williams {\it et al.}~\cite{williams:2006} 
explicitly derived the conversion efficiency 
as a function of the initial peak phase space density 
within the classical gas approximation. 
Both results are in good agreement with the experimental result~\cite{hodby:2005}. 
In this section, we explicitly derive the conversion efficiencies 
for the heteronuclear Feshbach molecule 
in population imbalanced case 
as a function of 
initial peak phase space density above $T_{\rm c}$, 
treating a majority component 
as a quantum degenerate gas. 

We use the Bose or Fermi distribution function for a majority component, 
while we use the Maxwell-Boltzmann distributions 
for a minority atomic component and heteronuclear molecules, 
because these minority densities are so small 
that interparticle distances are larger than the thermal de Broglie wavelength. 
In this approximation, 
the Bose and Fermi integrals for the minority component and heteronuclear molecules 
are given approximately by 
\begin{eqnarray}
{\mathcal G}_{n}(z) \approx z, &&{\mathcal F}_{n}(z) \approx z. 
\end{eqnarray} 

First, we consider the case with majority bosonic atoms within the approximation mentioned above, 
the number of each component is given by 
\begin{eqnarray}
N_{>}=
\left ( \frac{k_{\rm B}T}{\hbar\bar{\omega}_{>}}\right )^{3}
{\mathcal G}_{3}(z_{>}),\quad
N_{<}=
\left ( \frac{k_{\rm B}T}{\hbar\bar{\omega}_{<}}\right )^{3}
z_{<}, \quad
N_{\rm m}=
\left ( \frac{k_{\rm B}T}{\hbar\bar{\omega}_{\rm m}}\right )^{3}
z_{\rm m}. 
\end{eqnarray}

The entropy of each component is given by 
\begin{eqnarray}
\left \{
\begin{array}{ccl}
S_{>}&=&
k_{\rm B}N_{>}
\left [
4
\displaystyle{
\frac{{\mathcal G}_{4}(z_{>})}
{{\mathcal G}_{3}(z_{>})}
}
-\ln{z_{>}}\right ], 
\\
S_{<}&=&
k_{\rm B}N_{<}\left(4-\ln{z_{<}}\right), 
\\
S_{\rm m}&=&
k_{\rm B}N_{\rm m}\left(4-\ln{z_{\rm m}}\right). 
\end{array}
\right .
\end{eqnarray}

The molecular conversion efficiency is then given by 
\begin{eqnarray}
\chi_{0} = \frac{N_{\rm m}(\delta = 0)}{N_{\rm tot}}
\frac{1+\alpha}{\alpha}
=
\frac{1+\alpha}{\alpha}t^{3}\gamma_{\rm m}^{3}z_{\rm m}(\delta = 0), 
\end{eqnarray}
where $t \equiv k_{\rm B}T/(\hbar\bar{\omega}_{>}N_{\rm tot}^{1/3})$. 

According to the above equation and two constraints in Eqs. (\ref{eq2}) and (\ref{eq4}) 
for populations, 
we obtain $z_{\rm m}$, $z_{<}$, and $t^{3}{\mathcal G}_{3}(z_{>})$ 
at zero detuning $\delta = 0$ 
as a function of $\chi_{0}$; 
\begin{eqnarray}
\left \{
\begin{array}{ccl}
z_{\rm m}(\delta = 0)&=&
\displaystyle{
\frac{1}{\gamma_{\rm m}^{3}}\frac{1}{t^{3}}
}
\displaystyle{
\frac{\alpha}{1+\alpha}\chi_{0}
}, 
\\
\\
z_{<}(\delta = 0)&=&
\displaystyle{
\frac{1}{t^{3}}
}
\displaystyle{
\frac{1}{\gamma_{<}^{3}}\frac{\alpha}{1+\alpha}
}
(1-\chi_{0}), 
\\
\\
t^{3}{\mathcal G}_{3}(z_{>}(\delta = 0))
&=&
\displaystyle{
\frac{\alpha}{1+\alpha}
}
\displaystyle{
\left ( \frac{1}{\alpha}-\chi_{0} \right )
}. 
\end{array}
\right.
\end{eqnarray}

From the relation $z_{\rm m}=z_{>}z_{<}$ for fugacities 
at zero detuning $\delta=0$, 
the fugacity of the majority component $z_{>}$ as a function of $\chi_{0}$ 
is given by 
\begin{eqnarray}
z_{>}=\frac{z_{\rm m}}{z_{<}}
=
\frac{\chi_{0}}{1-\chi_{0}}
\left ( \frac{\gamma_{<}^{3}}{\gamma_{\rm m}^{3}} \right ) . 
\label{FugacityMajor}
\end{eqnarray}
Using these equations, 
we obtain the final total entropy at zero detuning $\delta = 0$ 
as a function of $\alpha$, $\chi_{0}$, $\gamma_{<}$ and $\gamma_{\rm m}$, 
\begin{eqnarray}
S_{\rm f}&=&
S_{>, {\rm f}} + S_{\rm m, f} + S_{<, {\rm f}}
\nonumber
\\
&=&
k_{\rm B}N_{\rm tot}
\left [ 
4\frac{1}{1+\alpha}
\left \{
\alpha
+ (1 - \alpha \chi_{0} )
\frac{{\mathcal G}_{4}
\left [ \left ( \frac{\gamma_{<}}{\gamma_{\rm m}} \right )^{3}  
\frac{\chi_{0}}{1-\chi_{0}}\right ]}
{{\mathcal G}_{3}
\left [ 
\left ( \frac{\gamma_{<}}{\gamma_{\rm m}} \right )^{3}  
\frac{\chi_{0}}{1-\chi_{0}} \right ]}
\right \} 
\right . 
\nonumber
\\
&&\quad
\left . 
-\ln{
\left (
\left \{
\frac{\alpha}{\gamma_{<}^{3}}
\frac{1-\chi_{0}}{1-\alpha\chi_{0}}
{\mathcal G}_{3}
\left[
\left ( \frac{\gamma_{<}}{\gamma_{\rm m}} \right )^{3}
\frac{\chi_{0}}{1-\chi_{0}}\right]
\right \}^{\frac{\alpha}{1+\alpha}}
\left(
\frac{\gamma_{<}}{\gamma_{\rm m}}
\frac{\chi_{0}}{1-\chi_{0}}
\right)^{\frac{1}{1+\alpha}}
\right )}
\right ].
\end{eqnarray}

On the other hand, 
the initial numbers of particles are given by 
\begin{eqnarray}
\left \{
\begin{array}{rcl}
N_{>,{\rm ini}}& =& 
\displaystyle{
\frac{1}{1+\alpha}N_{\rm tot}
}, 
\\
\\
N_{<,{\rm ini}}& =& 
\displaystyle{
\frac{\alpha}{1+\alpha}N_{\rm tot}
}. 
\end{array}
\right.
\end{eqnarray} 
The peak phase space density of a bosonic majority component is given by 
\begin{eqnarray}
\rho_{\rm pk}^{>} = 
\lambda_{>}^{3} n_{>}({\bf r} = {\bf 0})
= {\mathcal G}_{3/2}(z_{>}), 
\end{eqnarray}
where $\lambda_{>}$ is a thermal de Broglie wavelength 
of a majority component given by 
$\lambda_{>}=\left [ 2\pi\hbar^{2}/(m_{>}k_{B}T) \right]^{1/2}$, 
with $m_{>}$ being an atomic mass of the majority component. 
We obtain the fugacity $z_{>}$ 
as a function of $\rho_{\rm pk}^{>}$; 
\begin{eqnarray}
z_{>} = {\mathcal G}_{3/2}^{-1}(\rho_{\rm pk}^{>}). 
\end{eqnarray}

By making use of the equation of the second constraint 
$0 = N_{<,{\rm ini}}-\alpha N_{>,{\rm ini}}$, 
we obtain the relation between $z_{<,{\rm ini}}$ and $z_{>,{\rm ini}}$;  
\begin{eqnarray}
\gamma_{<}^{3}z_{<,{\rm ini}}=\alpha{\mathcal G}_{3}(z_{>,{\rm ini}}). 
\end{eqnarray}

The initial total entropy is given 
as a function of the ratio $\alpha$ and 
the initial peak phase space density of a majority component $\rho_{\rm pk, ini}^{>}$; 
\begin{eqnarray}
S_{\rm ini}&=&S_{<,{\rm ini}}+S_{>,{\rm ini}}
\nonumber
\\
&=&
k_{\rm B}N_{\rm tot}
\left [ 
4\frac{1}{1+\alpha}
\left \{ 
\alpha+
\frac{{\mathcal G}_{4}
\left [ {\mathcal G}_{3/2}^{-1}(\rho_{\rm pk, ini}^{>}) \right ] }
{{\mathcal G}_{3}
\left [ {\mathcal G}_{3/2}^{-1}(\rho_{\rm pk, ini}^{>}) \right ]}
\right \}
\right . 
\nonumber
\\
&&\quad
\left . 
-\ln{
\left (
\left \{
\alpha{\mathcal G}_{3}
\left [ {\mathcal G}_{3/2}^{-1}(\rho_{\rm pk, ini}^{>}) \right ]
\right \} ^{\frac{\alpha}{1+\alpha}}
\left [ {\mathcal G}_{3/2}^{-1}(\rho_{\rm pk, ini}^{>}) \right ]
^{\frac{1}{1+\alpha}}
\right ) }
\right ]. 
\end{eqnarray}
By equating the final entropy with the initial entropy, 
we obtain the relation between the molecular conversion efficiency $\chi_{0}$ 
and 
the initial peak phase space density of a majority component $\rho_{\rm pk, ini}^{>}$ as 
\begin{eqnarray}
&&
4(1-\alpha\chi_{0})
\frac{
{\mathcal G}_{4}
\left [ \left ( \frac{\bar{\omega}_{\rm m}}
{\bar{\omega}_{<}} \right )^{3} \frac{\chi_{0}}{1-\chi_{0}} \right ]
}
{
{\mathcal G}_{3}
\left [ \left ( \frac{\bar{\omega}_{\rm m}}
{\bar{\omega}_{<}} \right )^{3} \frac{\chi_{0}}{1-\chi_{0}} \right ]
}
- \ln{
\left (
\left \{
\frac{1-\chi_{0}}{1-\alpha\chi_{0}}
{\mathcal G}_{3}
\left [ 
\left (  \frac{\bar{\omega}_{\rm m}}{\bar{\omega}_{<}} \right)^{3}
\frac{\chi_{0}}{1-\chi_{0}}
\right]
\right \}^{\alpha}
\left ( 
\frac{\bar{\omega}_{\rm m}}{\bar{\omega}_{<}}
\right )^{3}
\frac{\chi_{0}}{1-\chi_{0}}
\right) 
}
\nonumber
\\
&=&4\frac{
{\mathcal G}_{4} [ 
{\mathcal G}_{3/2}^{-1}(\rho_{\rm pk, ini}^{>}) ]
}
{
{\mathcal G}_{4} [ 
{\mathcal G}_{3/2}^{-1}(\rho_{\rm pk, ini}^{>}) ]
}
-\ln{ 
\left ( 
\left \{
{\mathcal G}_{3}
\left [
{\mathcal G}_{3/2}^{-1}
(\rho_{\rm pk, ini}^{>})
\right ]
\right \}^{\alpha}
\left [
{\mathcal G}_{3/2}^{-1}(\rho_{\rm pk, ini}^{>})
\right ]
\right )}
\label{BoseChiRhoi}
\end{eqnarray}

Since the fugacity of the majority component $z_{>}$ should be less than unity, 
it follows from Eq. (\ref{FugacityMajor}) that 
the range of the conversion efficiency above the transition temperature of the majority atomic component BEC is 
limited as 
\begin{eqnarray}
\chi_{0} \leq \chi_{0,{\rm max}} \equiv 
\frac{\bar{\omega}_{<}^{3}}{\bar{\omega}_{<}^{3}+\bar{\omega}_{\rm m}^{3}}. 
\end{eqnarray}
This maximum conversion efficiency is consistent with our results discussed 
in Sec~\ref{SecBBB} and~\ref{SecBFF}, 
in the case where the majority component is bosonic.

Using the similar procedure, 
we consider the case with a majority fermionic component. 
In this case, a peak phase space density $\rho_{\rm pk, ini}^{>}$ is given by 
\begin{eqnarray}
\rho_{\rm pk, ini}^{>} = 
\lambda_{>}^{3} n_{>}({\bf r} = {\bf 0})
= {\mathcal F}_{3/2}(z_{>}). 
\end{eqnarray}
The relation analogous to Eq. (\ref{BoseChiRhoi}) for a fermionic majority component 
is obtained by simply replacing the Bose integral with the Fermi integral, 
\begin{eqnarray}
&&
4(1-\alpha\chi_{0})
\frac{
{\mathcal F}_{4}
\left [ \left ( \frac{\bar{\omega}_{\rm m}}
{\bar{\omega}_{<}} \right )^{3} \frac{\chi_{0}}{1-\chi_{0}} \right ]
}
{
{\mathcal F}_{3}
\left [ \left ( \frac{\bar{\omega}_{\rm m}}
{\bar{\omega}_{<}} \right )^{3} \frac{\chi_{0}}{1-\chi_{0}} \right ]
}
- \ln{
\left (
\left \{
\frac{1-\chi_{0}}{1-\alpha\chi_{0}}
{\mathcal G}_{3}
\left [ 
\left (  \frac{\bar{\omega}_{\rm m}}{\bar{\omega}_{<}} \right)^{3}
\frac{\chi_{0}}{1-\chi_{0}}
\right]
\right \}^{\alpha}
\left ( 
\frac{\bar{\omega}_{\rm m}}{\bar{\omega}_{<}}
\right )^{3}
\frac{\chi_{0}}{1-\chi_{0}}
\right) 
}
\nonumber
\\
&=&4\frac{
{\mathcal F}_{4} [ 
{\mathcal F}_{3/2}^{-1}(\rho_{\rm pk, ini}^{>}) ]
}
{
{\mathcal F}_{4} [ 
{\mathcal F}_{3/2}^{-1}(\rho_{\rm pk, ini}^{>}) ]
}
-\ln{ 
\left ( 
\left \{
{\mathcal F}_{3}
\left [
{\mathcal F}_{3/2}^{-1}
(\rho_{\rm pk, ini}^{>})
\right ]
\right \}^{\alpha}
\left [
{\mathcal F}_{3/2}^{-1}(\rho_{\rm pk, ini}^{>})
\right ]
\right )}
\label{FermiChiRhoi}
\end{eqnarray}

According to Eqs. (\ref{BoseChiRhoi}) and (\ref{FermiChiRhoi}), 
we find that 
the initial atomic population ratio $\alpha$ 
and the ratio of the trap frequencies of the minority atomic component 
and hetoronuclear molecular component 
$\bar{\omega}_{\rm m}/\bar{\omega}_{<}$ 
as well as 
the initial peak phase space density of the majority atomic component 
$\rho_{\rm pk, ini}^{>}$ 
determine  
the molecular conversion efficiency $\chi_{0}$ in the adiabatic ramp to form the heteronuclear Feshbach molecule.

In the high temperature region, 
one can also use the Maxwell-Boltzmann distribution for a majority component; 
${\mathcal G}_{n}(z_{>}) \approx z_{>}$ 
or ${\mathcal F}_{n}(z_{>}) \approx z_{>}$. 
In this situation, 
a fugacity of a majority component is approximately given by 
\begin{eqnarray}
z_{>} \approx \rho_{\rm pk, ini}^{>}.  
\end{eqnarray}
Quantum statistics is not important in this region. 
For classical gases, 
the relation between 
the molecular conversion efficiency $\chi_{0}$ 
and 
the initial peak phase space density 
of a majority component $\rho_{\rm pk, ini}^{>}$ 
is given by 
\begin{eqnarray}
4\frac{\alpha}{1+\alpha}\chi_{0}
+\ln{
\left\{
\left ( 
\frac{1-\chi_{0}}{1-\alpha\chi_{0}}
\right ) ^{\frac{\alpha}{1+\alpha}}
\left ( \frac{\bar{\omega}_{\rm m}}{\bar{\omega}_{<}} \right )^{3} 
\frac{\chi_{0}}{1-\chi_{0}}
\right \} }
= 
\ln{\rho_{\rm pk, ini}^{>}}. 
\label{ClassicalChiRhoi}
\end{eqnarray}
If we consider the population balanced case $\alpha = 1$ 
and equal trap frequencies $\bar{\omega}_{<} = \bar{\omega}_{\rm m}$, 
we recover the result given by Williams {\it et al}~\cite{williams:2006}; 
\begin{eqnarray}
\ln{\rho_{\rm pk, ini}^{>}}
&=&
2\chi_{0}
+\ln{
\left(
\frac{\chi_{0}}{1-\chi_{0}}
\right)}. 
\end{eqnarray}
By using the relation 
$\gamma_{<}^{3}\rho_{\rm pk, ini}^{<}=\alpha\rho_{\rm pk, ini}^{>}$ 
obtained from Eq. (\ref{eq4}) for the initial state, 
we also obtain the relation between 
the molecular conversion efficiency $\chi_{0}$ 
and 
the initial peak phase space density of a minority component $\rho_{\rm pk, ini}^{<}$; 
\begin{eqnarray}
4\frac{\alpha}{1+\alpha}\chi_{0}
+\ln{
\left\{
\left ( 
\frac{1-\chi_{0}}{1-\alpha\chi_{0}}
\right ) ^{\frac{\alpha}{1+\alpha}}
\left ( \frac{\bar{\omega}_{\rm m}}{\bar{\omega}_{<}} \right )^{3} 
\frac{\chi_{0}}{1-\chi_{0}}
\right \}}
= 
\ln{
\left[ \frac{1}{\alpha}
\left (  \frac{\bar{\omega}_{>}}
{\bar{\omega}_{<}}\right )^{3}
 \rho_{\rm pk, ini}^{<}\right]}.
\end{eqnarray}

\begin{figure}
\begin{center}
\includegraphics[width=16cm,height=6cm,keepaspectratio,clip]{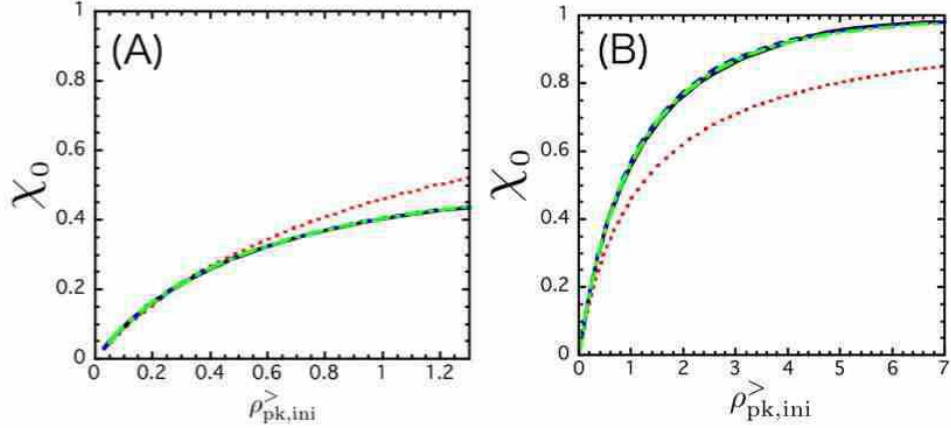}
\end{center}
\caption{
(Color online) 
The molecular conversion efficiency as a function of the initial peak phase space density of a majority component. 
We use the parameter $\alpha = 2/15$, and equal trap frequencies. 
(A) The solid and the dashed lines 
are the numerical results of $\{{\rm B}_{>} + {\rm B}_{<} \leftrightarrow {\rm B}_{\rm m}\}$ 
and  $\{{\rm B}_{>} + {\rm F}_{<} \leftrightarrow {\rm F}_{\rm m}\}$. 
The dot-dashed line is the result of our formula 
given in Eq. (\ref{BoseChiRhoi}). 
Those lines agree quite well with each other. 
The dotted line is the result of the classical limit 
given in Eq. (\ref{ClassicalChiRhoi}). 
(B) The solid and the dashed lines 
are the numerical results of $\{{\rm F}_{>} + {\rm B}_{<} \leftrightarrow {\rm F}_{\rm m}\}$ 
and  $\{{\rm F}_{>} + {\rm F}_{<} \leftrightarrow {\rm B}_{\rm m}\}$. 
The dot-dashed line is the result of our formula 
given in Eq. (\ref{FermiChiRhoi}). 
Those lines agree quite well with each other. 
The dotted line is the result of the classical limit 
given in Eq. (\ref{ClassicalChiRhoi}). 
}
\label{ChiRhoi.fig}
\end{figure}

Fig.~\ref{ChiRhoi.fig} shows the molecular conversion efficiency 
$\chi_{0}$ as a function of the initial peak phase space density 
$\rho_{\rm pk, ini}^{>}$. 
We assume $\alpha = 2/15$, and equal trap frequencies 
$\bar{\omega}_{>} = \bar{\omega}_{<} = \bar{\omega}_{\rm m}$. 
Fig.~\ref{ChiRhoi.fig} (A) shows the case where the majority atomic component is bosonic. 
The solid line is the numerical result 
of $\{{\rm B}_{>} + {\rm B}_{<} \leftrightarrow {\rm B}_{\rm m}\}$ 
discussed in Sec.~\ref{SecBBB}. 
The dashed line is the numerical result 
of $\{{\rm B}_{>} + {\rm F}_{<} \leftrightarrow {\rm F}_{\rm m}\}$ 
discussed in Sec.~\ref{SecBFF}. 
The dot-dashed line is the result of our formula 
given in Eq. (\ref{BoseChiRhoi}), 
reproducing two numerical lines quite well. 
The dotted line is the result of the classical limit 
given in Eq. (\ref{ClassicalChiRhoi}), 
which overlaps with three lines 
in the low initial peak phase space density limit. 
Fig.~\ref{ChiRhoi.fig} (B) shows the case where the majority atomic component is fermionic. 
The solid line is the numerical result 
of $\{{\rm F}_{>} + {\rm B}_{<} \leftrightarrow {\rm F}_{\rm m}\}$ 
discussed in Sec.~\ref{SecFBF}. 
The dashed line is the numerical result 
of $\{{\rm F}_{>} + {\rm F}_{<} \leftrightarrow {\rm B}_{\rm m}\}$ 
discussed in Sec.~\ref{SecFFB}. 
The dot-dashed line is the result of our formula 
given in Eq. (\ref{FermiChiRhoi}), 
reproducing two numerical lines quite well. 
The dotted line is the result of the classical limit 
given in Eq. (\ref{ClassicalChiRhoi}), 
which agrees with three lines 
in the low initial peak phase space density limit. 

\begin{figure}
\begin{center}
\includegraphics[width=7cm,height=7cm,keepaspectratio,clip]{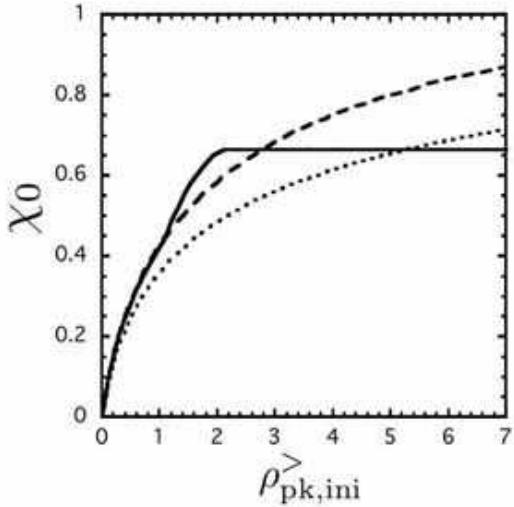}
\end{center}
\caption{
The molecular conversion efficiency as a function 
of the initial peak phase space density of a majority component, 
in the case $\{{\rm F}_{>} + {\rm B}_{<} \leftrightarrow {\rm F}_{\rm m}\}$ 
assuming $\alpha > \alpha_{\rm c}$. 
We assume $\alpha = 3/4 (> \alpha_{\rm c} = 1/2)$ and equal trap frequencies. 
The solid line is the numerical result. 
The dashed line is the result of our formula 
given in Eq. (\ref{FermiChiRhoi}). 
The dotted line is the result of the classical limit 
given in Eq. (\ref{ClassicalChiRhoi}). 
}
\label{ChiRhoiFFB.fig}
\end{figure}

Only in the case $\{{\rm F}_{>} + {\rm B}_{<} \leftrightarrow {\rm F}_{\rm m}\}$ 
discussed in Sec.~\ref{SecFBF}, 
the maximum conversion efficiency depends on 
the initial atomic population ratio $\alpha$. 
Fig.~\ref{ChiRhoiFFB.fig} shows the conversion efficiency 
as a function of the initial peak phase space density of 
the fermionic majority component 
in the case $\{{\rm F}_{>} + {\rm B}_{<} \leftrightarrow {\rm F}_{\rm m}\}$, 
assuming equal trap frequencies 
$\bar{\omega}_{>}=\bar{\omega}_{<}=\bar{\omega}_{\rm m}$ 
and $\alpha = 3/4 > \alpha_{\rm c} (= 1/2)$. 
In this case, the maximum conversion efficiency is given by $1/(2\alpha)$. 
The solid line is the numerical result 
discussed in Sec.~\ref{SecFBF}. 
The dashed line is the result of our formula 
given in Eq. (\ref{FermiChiRhoi}). 
The dotted line is the result of the classical limit 
given in Eq. (\ref{ClassicalChiRhoi}). 
According to Fig.~\ref{ChiRhoiFFB.fig}, 
our formula of Eq. (\ref{FermiChiRhoi}) 
does not reproduce the numerical result 
in the high initial peak phase space density region, 
differing from the case $\alpha = 2/15 (< \alpha_{\rm c} = 1/2)$ 
in Fig.~\ref{ChiRhoi.fig} (B). 
In the step deriving Eq. (\ref{FermiChiRhoi}), 
we treated that 
the minority atomic component and molecules as classical Maxwell-Boltzmann gases. 
For $\alpha > \alpha_{\rm c}$ ($= 1/2$), 
this approximation breaks down 
in the region where the quantum degeneracy 
of the minority atomic component cannot be ignored. 
The emergence of Bose statistics that the minority atomic component resides 
causes 
the disagreements between 
the numerical results and our formula as shown in Fig.~\ref{ChiRhoiFFB.fig}. 

To summarize, 
we found that the molecular conversion efficiency in an adiabatic sweep 
is well described as a function of the initial number ratio and the 
ratio of the trap frequencies of the minority atomic component and the heteronuclear molecular component 
as well as the initial peak phase space density 
\begin{eqnarray}
\chi_{0} = \chi_{0}\left ( 
\rho_{\rm pk, ini}^{>},\, \alpha, \,
\frac{\bar{\omega}_{<}}{\bar{\omega}_{\rm m}}
\right), 
\end{eqnarray}
in the case where populations of 
the minority component and the heteronuclear molecule are small.

\section{conclusion}
In this paper, 
we studied formations of heteronuclear Feshbach molecules 
in population imbalanced atomic gases, 
extending the recent work~\cite{williams:2006} on the
Feshbach molecule formation. 
At low temperature in quantum degenerate regime, 
quantum statistics of atoms plays an important role in determining conversion efficiencies. 

When the majority and minority atomic components are both bosonic, 
the maximum conversion efficiency is determined by the trap frequencies 
of the minority atomic component and heteronuclear molecules. 
Our calculation is in good agreement with the recent experiment~\cite{papp:2006} 
without any fitting parameters. 
An important finding is that 
one cannot convert any atoms 
into heteronuclear Feshbach molecules in the limit $T_{\rm ini} \rightarrow 0$. 
On the other hand, 
when gases are composed of fermionic atoms and bosonic atoms, 
the maximum conversion
efficiencies are determined by the trap frequencies 
of fermionic atoms and heteronuclear molecules as well as initial number ratio. 
In the case that both atomic components are fermionic,
the maximum conversion efficiency is $100\%$. 
In general, when {\it atoms} are not Bose condensed even at zero temperature at zero detuning, 
the molecular conversion efficiency reaches $100\%$. 
When one atomic component undergoes Bose-Einstain condensation 
but the other component does not at zero detuning, 
the conversion efficiency does not reach $100\%$, exhibiting a plateau.

In the region where the gases are not condensed, 
the conversion efficiency is described as an  explicit function of the initial number ratio of atoms 
and trap frequencies of minority component and heteronuclear molecule as well as the initial peak phase space density of a majority atomic component. 
We found that in the low-density region 
where Bose-Einstein condensation does not appear, 
the conversion efficiency is a monotonic function of the peak phase space density, 
but independent of statistics of the minority component.

Throughout this paper, we assumed equal trap frequencies 
in all figures for simplicity. 
This simple assumption is valid for loosely bounded Feshbach molecules 
composed of isotopic atoms. 
Although qualitative behaviors with respect to all figures 
do not change when we assume different trap frequencies, 
we note that 
the maximum conversion efficiencies depend on trap frequencies 
if BEC of the single atomic component appears.

Finally, 
we note that 
the theory of Feshbach molecule formation proposed by Williams {\it et al.}~\cite{williams:2006} 
brings the result quite different from the SPSS model of Ref.~\cite{hodby:2005}
in the low temperature region. 

\section{acknowledgment}
We are grateful to J. E. Williams for leading us to this problem. 
We thank S. B. Papp for helpful communications. 
S. W. thanks Y. Kato for useful comments.

\end{document}